\documentclass[a4paper,twocolumn,11pt,accepted=2022-01-11]{quantumarticle}
\pdfoutput=1
\usepackage[english]{babel}
\usepackage[utf8]{inputenc}
\usepackage[T1]{fontenc}
\usepackage{hyperref}
\usepackage{standalone}
\usepackage{tikz}
\usepackage{bm}
\usepackage{upquote}
\usepackage{listings}
\usepackage{caption}
\usepackage{subcaption}
\usepackage{graphicx}
\usepackage[normalem]{ulem}

\usepackage{hyperref}
\usepackage[numbers,sort&compress]{natbib}

\usepackage{listings}
\usepackage{amsmath,amsfonts,amsthm,bm}
\usepackage{physics}

\begin{document}
\title{Pulser: An open-source package for the design of pulse sequences in programmable neutral-atom arrays}
\author{Henrique Silvério}
\affiliation{Pasqal, 2 avenue Augustin Fresnel, 91120 Palaiseau, France}
\listcsgadd{author1affiliations}{\ref{fn:author}}
\author{Sebastián Grijalva}
\affiliation{Pasqal, 2 avenue Augustin Fresnel, 91120 Palaiseau, France}
\listcsgadd{author2affiliations}{\ref{fn:author}}
\author{Constantin Dalyac}
\affiliation{Pasqal, 2 avenue Augustin Fresnel, 91120 Palaiseau, France}
\author{Lucas Leclerc}
\affiliation{Pasqal, 2 avenue Augustin Fresnel, 91120 Palaiseau, France}
\author{Peter J. Karalekas}
\affiliation{Unitary Fund, Walnut, CA 91789, USA}
\orcid{0000-0001-6031-4800}
\thanks{(current address) AWS Center for Quantum Computing, Pasadena, CA 91125, USA}
\author{Nathan Shammah}
\affiliation{Unitary Fund, Walnut, CA 91789, USA}
\author{Mourad Beji}
\affiliation{Pasqal, 2 avenue Augustin Fresnel, 91120 Palaiseau, France}
\author{Louis-Paul Henry}
\affiliation{Pasqal, 2 avenue Augustin Fresnel, 91120 Palaiseau, France}
\author{Loïc Henriet}
\affiliation{Pasqal, 2 avenue Augustin Fresnel, 91120 Palaiseau, France}
\thanks{(corresponding author) loic@pasqal.io}
\maketitle
{\def\thefootnote{*}
    \footnotetext{\textsf{\label{fn:author}These authors contributed equally.}}}

\begin{abstract}
Programmable arrays of hundreds of Rydberg atoms have recently enabled the exploration of remarkable phenomena in many-body quantum physics. In addition, the development of high-fidelity quantum gates are making them promising architectures for the implementation of quantum circuits. 

We present here \textit{Pulser}, an open-source Python library for programming neutral-atom devices at the pulse level. The low-level nature of Pulser makes it a versatile framework for quantum control both in the digital and analog settings. The library also contains simulation routines for studying and exploring the outcome of pulse sequences for small systems.
\end{abstract}

\section{Introduction}
Individual neutral atoms trapped in optical tweezers\,\cite{Saffman10,Saffman2016,barredo_atom-by-atom_2016,endres_atom-by-atom_2016,barredo_synthetic_2018,Browaeys20,Henriet2020quantum,Morgado20,Beterov20,Wu21} have emerged as a powerful platform for quantum information processing. In those systems, one uses the strong interaction between atoms excited to a \emph{Rydberg state} to generate entanglement. This interaction can be exploited for various tasks, including the creation of fast and robust quantum gates\,\cite{Jaksch00,Isenhower10,levine_high-fidelity_2018,Levine19} for digital quantum information processing, the generation of tunable spin models for the study of many-body quantum physics\,\cite{labuhn2016tunable,Bernien17,Lienhard18,leseleuc_observation_2019,omran_generation_2019}, or the deterministic creation of large highly entangled GHZ states with up to 20 atoms\,\cite{omran_generation_2019}.

Quantum computing and quantum simulation schemes for neutral-atom Quantum Processing Units (QPUs) are extremely flexible, in part due to the fact that, unlike for most other platforms, the QPU (and therefore its qubit connectivity) is not hard-wired during the manufacturing process. In fact, in neutral-atom optical traps, the connectivity of the QPU can be reprogrammed at \textit{every single run}. This opportunity opens up new avenues for exploration on these devices but also requires providing the user with a lower-level description of the physical setup.

We here introduce \emph{Pulser}~\cite{Pulser}, an open-source software library for designing pulse sequences for neutral-atom QPUs. The main goal of this library is to serve as an interface between experienced users and neutral-atom quantum hardware. Using Pulser, users can control all the relevant physical parameters of a pulse-level quantum program. This program can subsequently be sent to QPUs via external tools and executed on the hardware after a device-specific compilation step. Providing users with a pulse-level access to QPUs will allow them to explore advanced pulse-shaping techniques facilitating the design of new quantum gate protocols, the implementation of optimal control and error mitigation techniques, or the exploration of mixed digital-analog algorithms\,\cite{Parra-rodriguez_20}. Exposing the low-level details of the hardware will also enable algorithm developers to design software procedures while keeping the particularities of the hardware in mind. Besides providing an interface to a QPU, Pulser includes a built-in emulator that faithfully reproduces the hardware behavior. This is intended to make the design of experiments easier for both the theoretician and the experimentalist.

Pulser is an open-source Python library licensed under the Apache License 2.0 and intended for the community of quantum scientists and engineers working on neutral-atom devices. The development and use of open-source software has enabled the flexible and efficient simulation of quantum systems \cite{Johansson_2012,Johansson_2013}, the connection of quantum processors to the cloud \cite{Zeng_2017_Nature, Karalekas_2020} and their use to launch jobs running quantum hardware. These tasks appear as quantum circuits expressed in intermediate representation, or in the more recent framework of hybrid classical-quantum algorithms, which require a fast and tailored interoperability between quantum and classical computing units with classical optimization routines \cite{McClean_2016_NJP,Bharti_2021,Endo_2021}. More recently, specific tools and frameworks have also put more focus on the pulse-level description of quantum controls \cite{Alexander_2020_QST,Ball_2020,Wittler_2021, QUA-libs, li2021qutip-qip}, and their optimization to mitigate the impact of noise on the coherence of the intended protocols \cite{Goerz_2019_SciPost} in the so-called era of noisy-intermediate scale quantum computing (NISQ) \cite{Preskill_2018_Quantum}. However, no existing open-source tool has yet specialized its features to the physics of neutral-atom quantum processors, which, due to the peculiarities of their physical properties, provide largely untapped opportunities both for quantum simulation and quantum computing tasks.

In this paper, we start by briefly reviewing in Section \ref{section:description} the main physical ingredients that determine the dynamics of Rydberg atom arrays. We then present in Section \ref{section:sequences} a guided walk-through for designing pulse sequences with Pulser, followed by examples on how Pulser can be leveraged for some notable use-cases in Section \ref{section:usecases}. Finally, we conclude and comment on the future directions and extensions for this tool.

\section{Description of the physical system}
\label{section:description}
In neutral-atom devices, atoms are trapped by arrays of optical tweezers\,\cite{Nogrette14} in arbitrary, customizable patterns. In Pulser, we call this ensemble of atoms and their specific positions the \texttt{Register}. For each atom in a \texttt{Register}, the quantum information is encoded in specific electronic energy levels, which are reached through optical addressing techniques. In Pulser, the components responsible for driving these transitions are called \texttt{Channel}s and they do so through the emission of \texttt{Pulse}s.

\subsection{Driving two-level transitions}

We define a pulse as the modulation of a channel's output amplitude, detuning, and phase over a finite duration $\tau$. For a channel targeting the transition between energy levels $a$ and $b$, with resonance frequency $\omega_{ab} = |E_{a}-E_{b}|/\hbar$, the output amplitude determines the \textit{Rabi frequency}\footnote{The Rabi frequency determines the amplitude of a signal (in frequency units), not the frequency itself.} $\Omega(t)$, and the detuning $\delta(t)$ is defined relatively to $\omega_{ab}$ and the frequency of the channel's output signal $\omega(t)$, as $\delta(t) = \omega(t) - \omega_{ab}$. Additionally, the phase $\varphi$ of a pulse can be set to an arbitrary, constant value.

\begin{figure}[h]
\centering
\includegraphics[width=0.9\columnwidth, trim=0 0.5cm 0 0.8cm, clip]{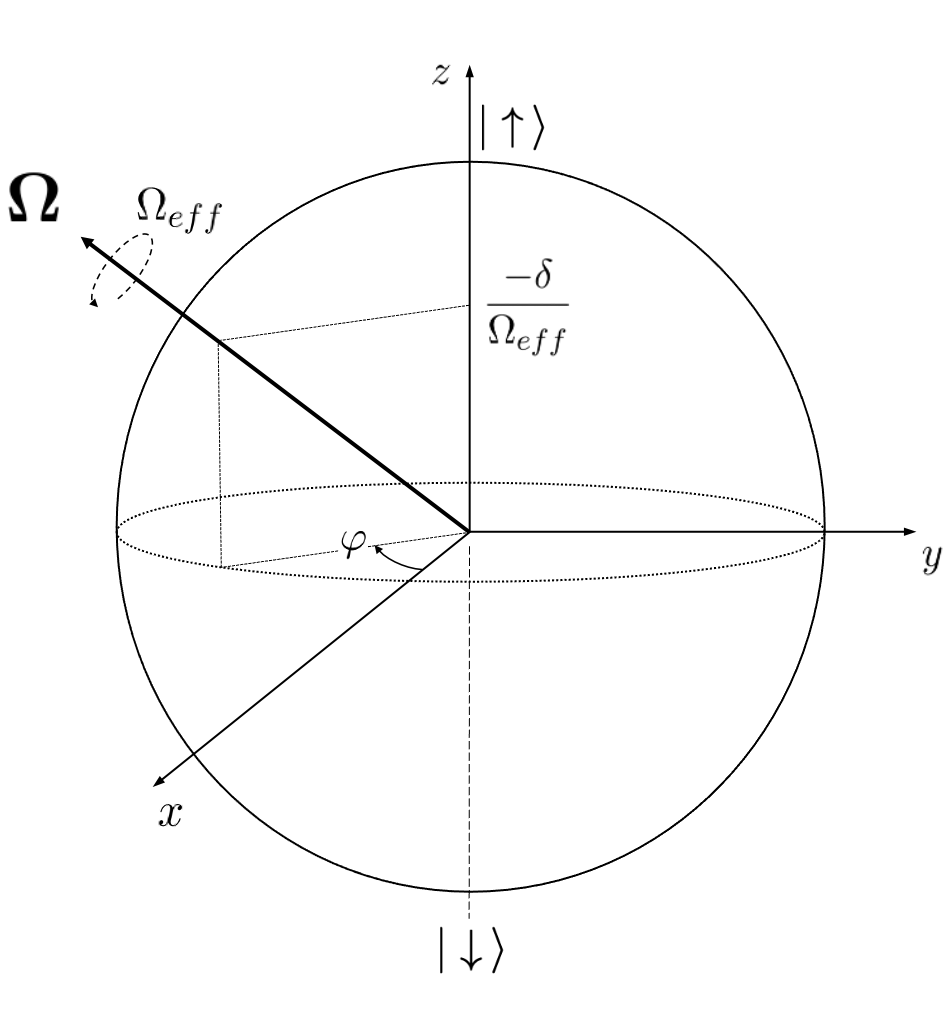}
\caption{Representation of the drive Hamiltonian's dynamics as a rotation in the Bloch sphere.}
\label{fig:bloch_rotation}
\end{figure}

A pulse-driven transition between two energy levels can be mapped to a spin-$1/2$ system through the drive Hamiltonian:
\begin{equation}\label{drive_ham}
H^D(t)= \frac{\hbar}{2} \mathbf{\Omega}(t)\cdot\bm{\sigma},
\end{equation}
where $\bm{\sigma} = (\sigma^x,\sigma^y,\sigma^z)^T$ is the Pauli vector and $\bm{\Omega}(t) =
(\Omega(t) \cos(\varphi),-\Omega(t) \sin(\varphi),-\delta(t))^T$ the rotation vector. Since there can be multiple transitions being addressed overall, we explicitly define the Pauli matrices for a transition between states $\ket{a}$ and $\ket{b}$ (where $E_b > E_a$) to be:
\begin{align}
    \sigma^x &= \op{a}{b} + \op{b}{a}\\
    \sigma^y &= i\op{a}{b} - i\op{b}{a} \\
    \sigma^z & = \op{b}{b} - \op{a}{a}
\end{align}
In the Bloch sphere representation, for each instant $t$, this Hamiltonian describes a rotation around the axis $\mathbf{\Omega}$ with angular velocity $\Omega_{eff} = |\bm{\Omega}| = \sqrt{\Omega^2 + \delta^2}$, as illustrated in Figure \ref{fig:bloch_rotation}.

\subsection{Rydberg states}
\label{sec:rydberg}

In neutral-atom devices, atoms are driven to Rydberg states as a way to make them interact over large distances. Depending on the electronic levels that are involved in the process, the atoms experience different types of interactions, translating into different Hamiltonians~\cite{Browaeys20}.\\

\begin{figure}[h]
\centering
\includegraphics[scale=0.4]{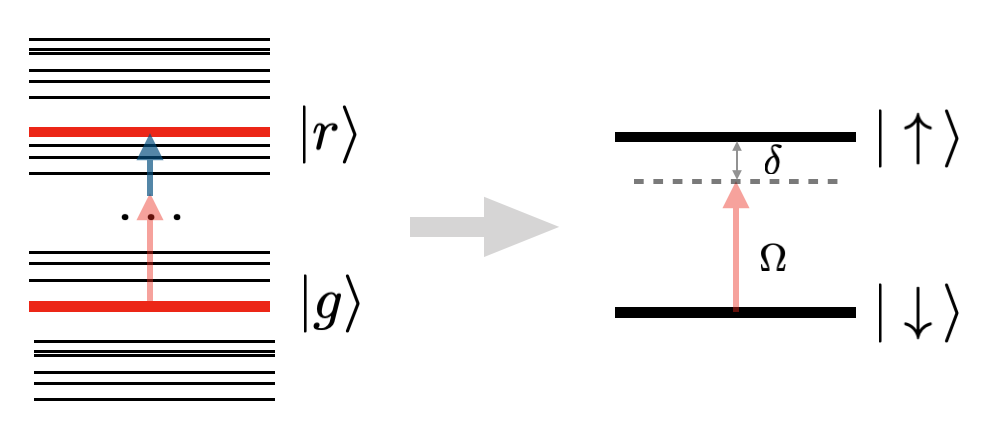}
\caption{The excitation between a ground and Rydberg state is taken as two-level transition with  Rabi frequency $\Omega$ and detuning $\delta$. In Pulser, it is called the \texttt{ground-rydberg} basis and is addressed by \texttt{Rydberg} channels.}
\label{Ising_config}
\end{figure}

In the so-called ``Ising'' configuration, which is obtained when the spin states are one of the ground states $\ket{g}$ and a Rydberg state $\ket{r}$~\cite{schauss2015crystallization,labuhn2016tunable,Bernien17,leseleuc2018accurate}  (the \texttt{ground-rydberg} basis in Pulser, see Fig. \ref{Ising_config}.), the interaction adds a term to the drive Hamiltonian describing the transition between those states: 

\begin{equation}
\mathcal H^{gr}(t) = \sum_i \left(H^D_i(t) + \sum_{j<i}\frac{C_6}{(R_{ij})^6}\hat n_i \hat n_j \right),
\label{eq:ising_ham}
\end{equation}

\noindent where $\hat n_i$ denotes the projector $\ketbra{r}{r}_i$ on the $i$-th atom, $R_{ij}$ is the distance between atoms $i$ and $j$, and $C_6$ is a constant depending on the specific Rydberg level $\ket{r}$.

\begin{figure}[h]
\centering
\includegraphics[width=\linewidth]{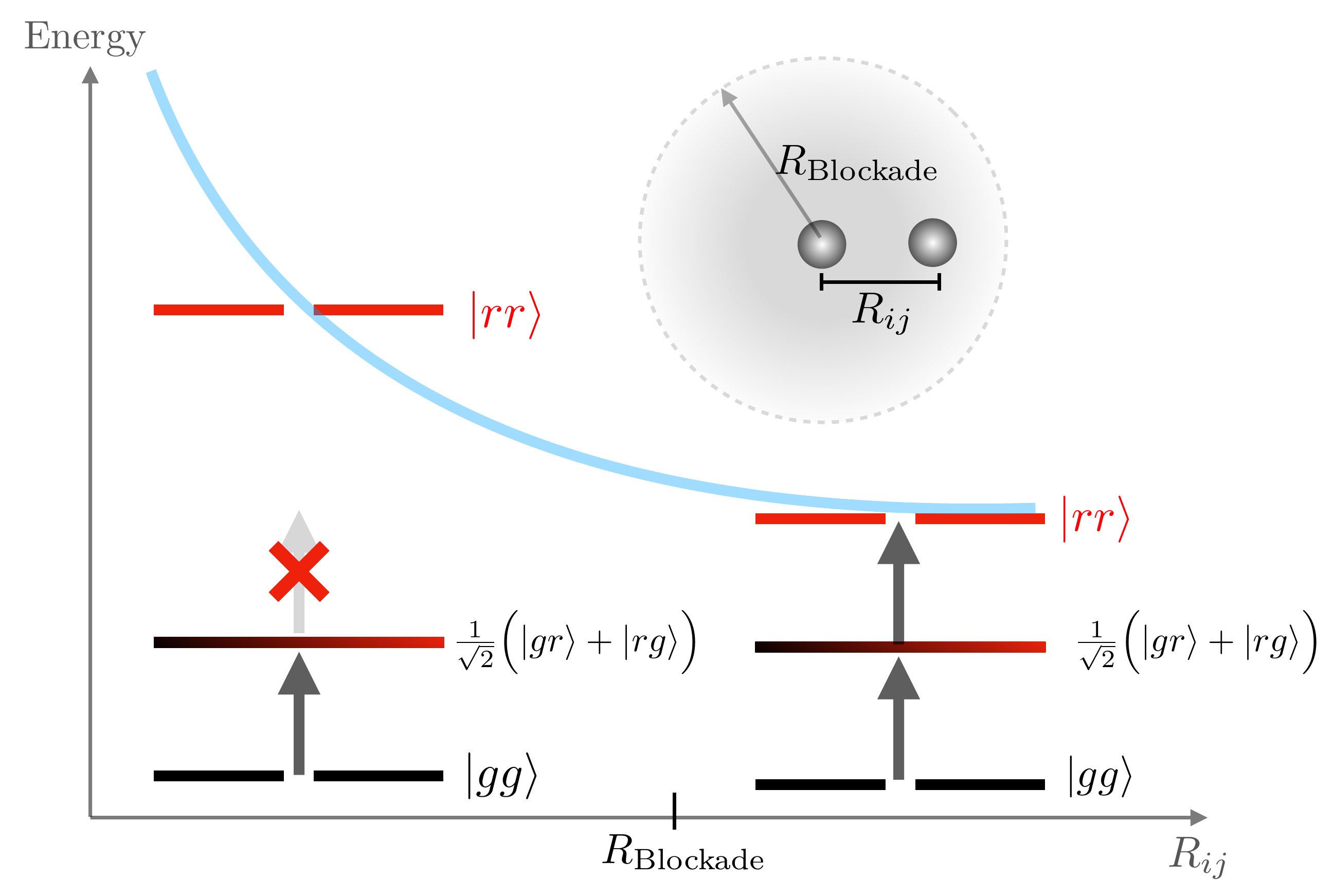}
\caption{Two nearby atoms interact via a van der Waals potential which shifts the energy level for the doubly excited state $|rr\rangle$. If the laser excites simultaneously both atoms, the lower-energy entangled state $(|gr\rangle + |rg\rangle)/\sqrt{2}$ is produced instead.}
\label{blockade}
\end{figure}

The effect of this interaction is expressed in the so-called \textit{Rydberg blockade}, illustrated in Figure \ref{blockade}. When driving two atoms from $\ket{g}$ to $\ket{r}$ with a resonant pulse ($\delta=0$) of Rabi frequency $\Omega$, the interaction prevents the simultaneous excitation of the two atoms in the state $ \ket{r}$ if $\hbar\Omega \ll C_6/R^6$. Instead, the system evolves between the $\ket{gg}$ and the entangled state $\ket{\psi_+}=(\ket{gr}+\ket{rg})/\sqrt{2}$\footnote{Note that the state $\ket{\psi_-}=(\ket{gr}-\ket{rg})/\sqrt{2}$ is not coupled to $\ket{gg}$ by the laser}, with an effective Rabi frequency of $\sqrt{2}\Omega$. Similarly, an atom's excitation from $\ket{g}$ to $\ket{r}$ is suppressed if there is another atom in the Rydberg state at a distance $R \ll R_b$, where 
\begin{equation}\label{eq:blockade_radius}
    R_b=\left(\frac{C_6}{\hbar\Omega}\right)^{1/6}
\end{equation}
is the \textit{Rydberg blockade radius}.

%



By acting on all atoms with the same global pulse, addressing the $\ket{g} \leftrightarrow \ket{r}$ transition with $\varphi=0$, one obtains the Hamiltonian:
\begin{equation}
    \mathcal H(t) = \sum_i \left(\frac{\hbar\Omega(t)}{2} \sigma_i^x 
    -\hbar \delta(t) \hat n_i +\sum_{j<i} \frac{C_6}{(R_{ij})^6}\hat n_i \hat n_j \right)
\label{eq:global_ising}
\end{equation}
Since $\hat n_i=(1+\sigma_i^z)/2$, Eq. (\ref{eq:global_ising}) represents a quantum Ising Hamiltonian with transverse field $\Omega(t)$ and Ising couplings $J_{ij} \propto 1/R_{ij}^6$.\\

Other configurations exist, but the corresponding channels are currently not implemented in Pulser. In the so-called ``XY'' configuration, which is obtained when the spin states $\ket{\downarrow}$ and $\ket{\uparrow}$ are two Rydberg states that are dipole-coupled, such as $\ket{nS}$ and $\ket{nP}$~\cite{barredo2015coherent,orioli2018relaxation,leseleuc_observation_2019,Scholl2021}, the dipolar interaction generates an exchange-like term of the form
\begin{equation}
\mathcal{H}_{int}=  2\sum_{i \neq j}\frac{C_3}{R_{ij}^3} \left(\sigma_i^x \sigma_j^x + \sigma_i^y \sigma_j^y \right),
\label{eq:XY_hamiltonian}
\end{equation}
where $C_3$ is a coefficient which depends on the selected Rydberg states. In this configuration, the driving is induced by a microwave field instead of an optical laser field.\\

The continuous manipulation of a system's Hamiltonian is referred to as the \textit{analog approach} to quantum computing. The very high level of control in this analog mode has several applications, the most direct one being the quantum simulation of interacting many-body systems. This includes observing the out-of-equilibrium dynamics of the system after a quench or a drive\,\cite{Bernien17,Bluvstein21}, or determining the ground-state of a specific Hamiltonian through adiabatic variation of its parameters\,\cite{Lienhard18,scholl2020programmable,ebadi2020quantum}. Additionally, one can use the blockade effect to reproduce a connectivity structure among the atoms, which can be exploited to solve combinatorial problems that can be mapped onto the Ising Hamiltonian\,\cite{pichler2018quantum,henriet2020robustness, dalyac2020qualifying}. Those applications will be studied in more detail in Section \ref{section:usecases}.\\

\subsection{Digital addressing and quantum gates}

In opposition to the analog approach stands the \textit{digital approach}, in which a system's state evolves through a series of discrete manipulations of its qubits' states, known as quantum gates. This is the underlying approach in quantum circuits and can be replicated on neutral-atom devices at the pulse-level. To achieve this, the qubit states are encoded in two hyperfine ground states of the system, named \textit{ground}, $\ket{g}\equiv\ket{0}$, and \textit{hyperfine}, $\ket{h}\equiv\ket{1}$. In Pulser, these states form the \texttt{digital} basis, which is addressed by \texttt{Raman} channels.

For these energy levels, the Rydberg blockade effect is not present, so the resulting dynamics from a pulse addressing the $\ket{h} \leftrightarrow \ket{g}$ transition of an atom will be dictated by the driving Hamiltonian of eq. (\ref{drive_ham}). For a pulse of duration $\tau$, the time evolution of this system is dictated by the operator
%
\begin{equation}
U(\bm \Omega, \tau) = T \exp \left[ -  \frac{i}{2} \int_0^\tau \bm \Omega(t) \cdot \bm \sigma \mathrm dt\right],
\end{equation}
where $T$ denotes the time-ordering operator. The unitary $U$ describes a rotation around the time-dependent axis $\bm\Omega(t)$. For a resonant pulse (i.e. one with $\delta = 0$) of phase $\varphi$, we get a rotation angle of
\begin{equation}\label{rot_angle}
    \theta = \int_0^\tau \Omega(t) \mathrm dt
\end{equation}
around the fixed axis 
\begin{equation}\label{rot_axis}
   \bm e(\varphi) = (\cos\varphi, -\sin\varphi, 0),
\end{equation}
situated on the equator of the Bloch sphere. The corresponding unitary operator is
\begin{equation}
\begin{split}\label{res_rotation}
    R_{\bm e(\varphi)}(\theta) &= e^{-i\frac{\theta}{2}(\cos(\varphi) \sigma_x - \sin(\varphi) \sigma_y)}\\
    &=  e^{i\frac{\varphi}{2}\sigma_z}e^{-i\frac{\theta}{2}\sigma_x}e^{-i\frac{\varphi}{2}\sigma_z}\\
    &= R_{ z}(-\varphi)R_{ x}(\theta)R_{ z}(\varphi)
\end{split}
\end{equation}
which, as its decomposition shows, can be thought of as a rotation around the Bloch sphere's $x$-axis, conjugated by $z$-rotations (i.e. phase gates). By following this gate with another $z$-rotation (which can be achieved virtually through a shift in the phase reference frame \cite{IBM_VZ-gates}), we can then construct any arbitrary single-qubit gate,
\begin{equation}
\begin{split}
    U(\gamma, \theta, \varphi) &= R_z(\gamma + \varphi) R_{\bm e(\varphi)}(\theta) \\
    &= R_z(\gamma)R_x(\theta)R_z(\varphi),
\end{split}
\end{equation}
which relies solely on resonant pulses and phase reference frame changes. 

\begin{figure}
    \centering
    \includegraphics[width=\columnwidth]{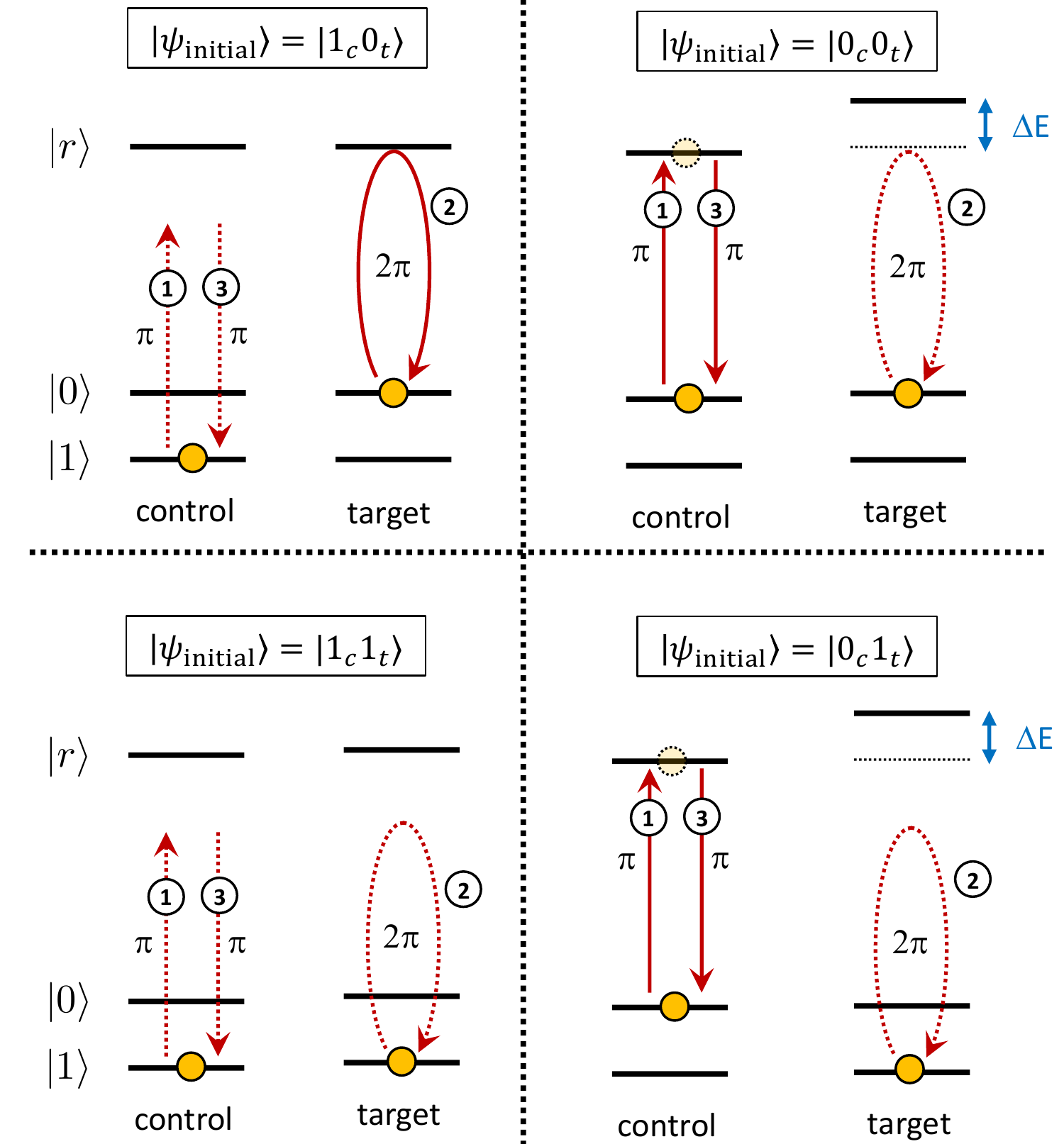}
    \caption{CZ implementation protocol, where three consecutive pulses address the $\ket{g} \leftrightarrow \ket{r}$ transition. Full (dashed) red lines illustrate if a given pulse is effective (not effective).}
    \label{fig:cz_protocol}
\end{figure}

With the ability to perform arbitrary single-qubit gates, all that is missing in order to have a universal quantum gate set is an appropriate two-qubit gate. In neutral-atom devices, the native gate is the controlled-Z (CZ); its implementation leverages the Rydberg blockade effect to flip the phase of a two-qubit subsystem, conditionally on its initial state. To achieve this, the atoms, depending on their initial state, are sequentially brought to an auxiliary Rydberg state and back by a sequence of three consecutive laser pulses. The effect of this pulse sequence on the four computational states $\ket{00}$, $\ket{01}$, $\ket{10}$ and $\ket{11}$ is shown in Figure \ref{fig:cz_protocol}. The first $\pi$-pulse attempts to bring the \textit{control} atom into the Rydberg state, which will condition whether the \textit{target} will be able to go to $\ket{r}$ and back with the second $2\pi$-pulse, due to the Rydberg blockade. In particular, the $\ket{00}$ state does not experience two consecutive phase flips as it would if the atoms did not interact (see top right panel of Fig. \ref{fig:cz_protocol}, where the Rydberg energy level of the target atom experiences a shift in energy $\Delta E$ due to the control atom being in $\ket{r}$). Finally, the third pulse brings the \textit{control} back to its original state, in case it was excited. In the end, all states but $\ket{11}$ experience a phase flip (see bottom left panel of Fig. \ref{fig:cz_protocol}), and the process corresponds to an application of a CZ within a global phase of $\pi$ (see table \ref{tab:cz}).

\begin{table}[h]
    \centering
    \begin{tabular}{c||c | c |c | c}
        Initial state & $\ket{00}$ & $\ket{01}$ & $\ket{10}$ & $\ket{11}$ \\
        \hline
        Final state &  $-\ket{00}$ & $-\ket{01}$ & $-\ket{10}$ & $\ket{11}$ \\
    \end{tabular}
    \caption{Effect of the CZ pulse sequence in the computational basis.}
    \label{tab:cz}
\end{table}

Altogether, the space in which the atom arrays can evolve is here composed of three states: two non-interacting hyperfine states, $\{ |g\rangle , |h\rangle \}$ and an auxiliary Rydberg state $|r\rangle $ which is only populated in the transient dynamics of the system during the application of multi-qubit gates. The Rydberg state thus plays both the role of an excited state in the \texttt{ground-rydberg} basis in the analog approach, and of an auxiliary state for realizing multi-qubit gates in the \texttt{digital} basis.

Since the most general sequence, with channels addressing both transitions, involves three energy levels, the size of the Hilbert space is $3^N$, where $N$ is the number of atoms in the array. Typically, a personal computer can efficiently simulate the evolution of systems of size below $N \sim 20$.

\subsection{Measurement}

In neutral-atom devices, the measurement process encompasses all the atoms in the register and necessarily terminates the computation. The process is based on the fluorescence imaging of the atom array, where ``bright'' and ``dark'' sites are linked to the measured states\,\cite{Fuhrmanek11}. Due to there being multiple addressable transitions, the final state of each atom can be in more than two levels and, therefore, the measurement's outcome will be basis-dependent. In Pulser, the convention is to associate the state that is exclusive to a given basis to $1$ and all other outcomes to $0$, as summarized in table \ref{tab:measures}.

\begin{table}[h]
    \centering
    \begin{tabular}{c||c | c}
        \cline{2-3}
        &  \lstinline[] $ground-rydberg$ & \lstinline[] $digital$\\ \hline
        $\ket{r}$ & 1 & 0 \\
        $\ket{g}$ & 0 & 0 \\
        $\ket{h}$ & 0 &  1 \\
    \end{tabular}
    \caption{Corresponding measurement outcome of each state, depending on the measurement basis.}
    \label{tab:measures}
\end{table}


\section{Designing sequences with Pulser}
\label{section:sequences}

Pulser is written in the Python language to facilitate adoption by a widespread audience. It is also Open Source Software and can be accessed and developed by any interested contributor. To install the latest stable version of the Pulser package, type into a terminal (where Python version 3.7 or greater is installed):
\begin{lstlisting}[numbers=none]
pip install pulser
\end{lstlisting}

\subsection{Code Architecture}

\begin{figure*}[t!]
\centering
\includegraphics[width=\linewidth]{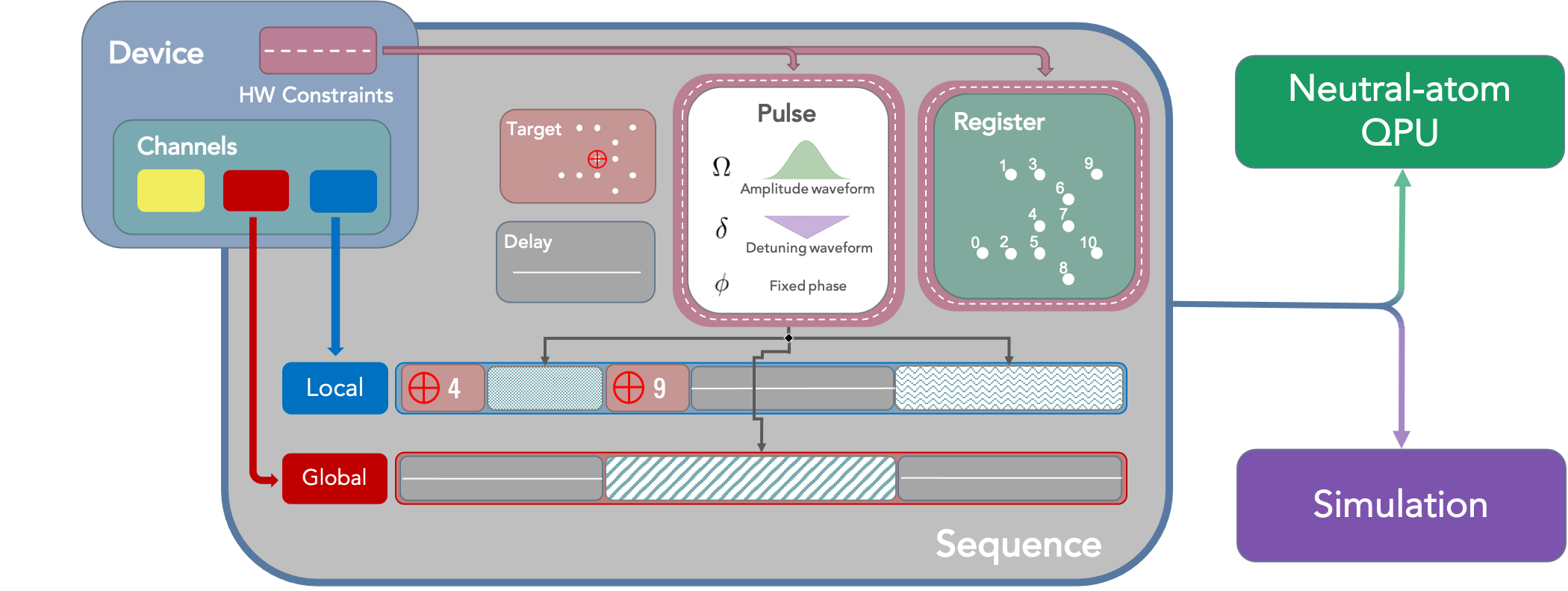}
\caption{Relationship between the main Pulser classes. The central object is the \texttt{Sequence}, which is linked to a \texttt{Device}. The \texttt{Device} holds the available \texttt{Channels} --- which are selected and declared in the \texttt{Sequence} --- and information of the hardware constraints. These constraints are enforced upon the \texttt{Register}, where the neutral-atom array is defined, and upon the \texttt{Pulses}. Each \texttt{Pulse}, defined by its amplitude and detuning \texttt{Waveforms} and a fixed phase, populates the declared channels alongside other commands like \texttt{target} --- which points local addressing channels to specific qubits --- and \texttt{delay} --- which idles the channel. The resulting \texttt{Sequence} can then be sent for execution on the neutral-atom QPU or emulated through Pulser's \texttt{Simulation} class.}
\label{fig:architecture}
\end{figure*}

The main components of Pulser are:

\begin{itemize}
    \item[(\emph{i})] The \texttt{Device}, which consists of a series of specifications that characterize the hardware, including the chosen Rydberg level, the ranges of the amplitudes and frequencies of the lasers, the minimal and maximal distance between the atoms, and the different channels that can be declared (see (\emph{iii}) below).
    \item[(\emph{ii})] The \texttt{Register} stores the information about the coordinates of the atoms and their respective ID's, which serve to identify them when targeting specific operations.
    \item[(\emph{iii})] The \texttt{Channel}s represent the action of the lasers and are organized by addressing (local or global) and the type of transition (Rydberg or Raman).
    \item[(\emph{iv})] \texttt{Waveform}s are the basic building blocks of a pulse. They can have custom or predetermined shapes, like a ramp waveform or a Blackman waveform, all of them indicating the specific profile of the waveform and its duration.
    \item[(\emph{v})] \texttt{Pulse}s consist of waveforms for the amplitude and the detuning. They can be further shifted by a phase. Once a \texttt{Pulse} is constructed it has to be added to the sequence indicating which atom(s) are targeted and what channel will implement it.
    \item[(\emph{vi})] The \texttt{Sequence} contains the schedule of the pulses in each channel. It is also linked with a \texttt{Register} and the \texttt{Device} in which it is to be executed. This is the information that can be sent to a real neutral-atom QPU or simulated on a classical computer.
    \item[(\emph{vii})] \texttt{Simulation} is included to emulate results for the application of sequences on small instances. In Pulser, we have made use of the QuTiP libraries\,\cite{Johansson_2012,Johansson_2013} for the simulation of quantum systems. Each simulation run returns a specialized object that holds the results and features methods for post-processing them.
\end{itemize}

\subsection{Pulse Sequence Creation: A guided walkthrough}

We will show, step-by-step, how a basic pulse sequence is created with Pulser by going through the paradigmatic protocol for creating the Bell state $\ket{\Phi^+} = \frac{1}{\sqrt{2}}(\ket{00} + \ket{11})$. 

\begin{figure}[h]
\centering
\includegraphics[width=1.\columnwidth]{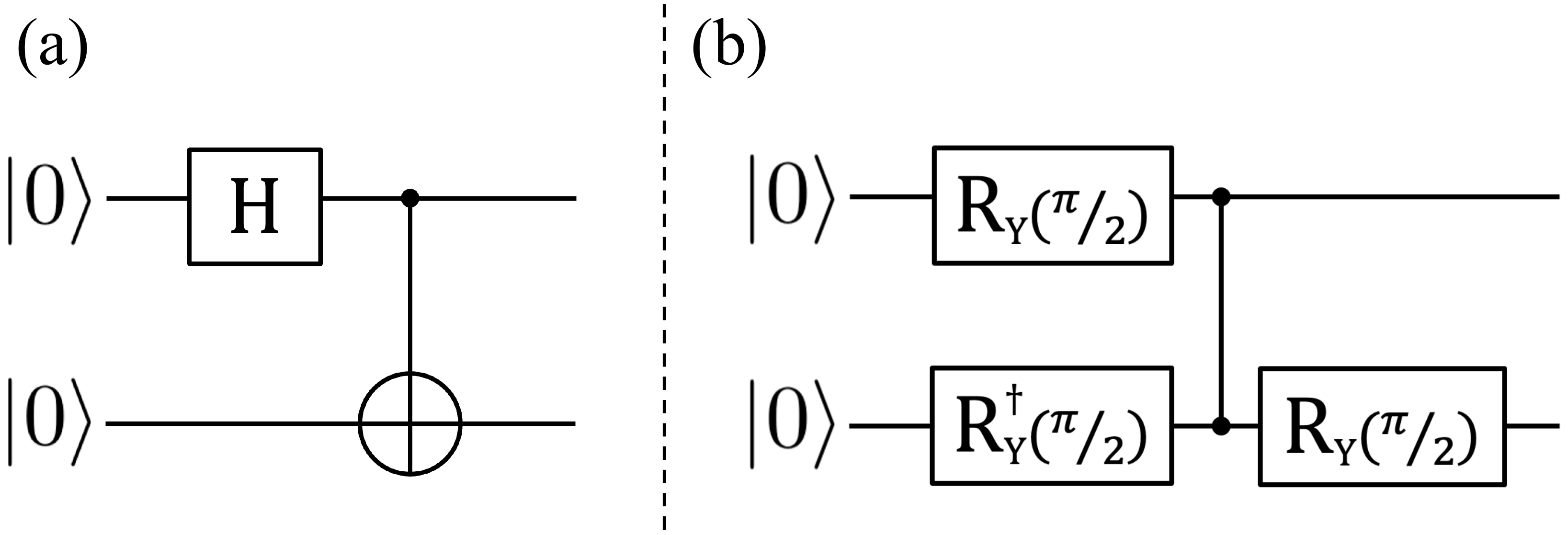}
\caption{ (a) Typical quantum circuit for generating the Bell state $\ket{\Phi^+}.$ (b) Equivalent circuit to that of Figure \ref{fig:bell_circuit}(a), adapted for execution on a neutral-atom device. }
\label{fig:bell_circuit}
\end{figure}

Starting from the archetypal circuit in Figure \ref{fig:bell_circuit}(a), we need to adapt it for execution on neutral-atom devices. The major and necessary change is the decomposition of the CNOT gate, which is not native in neutral-atom devices, to an equivalent construction based on the CZ gate. Additionally, we replace the Hadamard with an equivalent gate (when starting from $\ket{0}$) that can be executed with a resonant pulse, which reduces the number of free parameters in the pulse, making it less error-prone. These changes result in the equivalent circuit of Figure \ref{fig:bell_circuit}(b).

From this quantum circuit, we can create the corresponding pulse sequence in Pulser. For this effect, the following modules will be needed:

 \begin{lstlisting}[firstnumber=1]
 import numpy as np
 import pulser
 from pulser import Pulse, Sequence, Register
 \end{lstlisting}

\subsubsection{Creating the Register}

The \texttt{Register} defines the positions of the atoms and their respective names. There are multiple ways of defining a \texttt{Register}, the most customizable one being to create a dictionary that associates a name (the key) to a coordinate in micrometers (the value). In this case, we only need two qubits, so we will create a \texttt{Register} with two atoms, 4 $\mu$m apart, which we will name \lstinline{'c'} and \lstinline{'t'}.

\begin{lstlisting}[firstnumber=4]
# Coordinates in micrometers
qubits = {'c': (-2, 0), 't': (2, 0)}
reg = Register(qubits)
\end{lstlisting}

Note that, although we can arbitrarily specify the position of each qubit, the \texttt{Device} with which we will associate it will impose some restrictions, limiting in particular the minimal distance between any two qubits and the maximal distance of each qubit from the center of the array.





\subsubsection{Initializing the Sequence}

To create a \texttt{Sequence}, one has to provide it with the \texttt{Register} instance and the \texttt{Device} in which the sequence will be executed. Each \texttt{Device} object is associated with a physical device and holds its particular constraints. When linked to a \texttt{Sequence}, it dictates whether the chosen register is valid and, as the sequence is created, validates all instructions.
We import the device (in this case, \texttt{Chadoq2}, which is Pasqal's research and development prototype) from \texttt{pulser.devices} and initialize our sequence with the freshly created register:

\begin{lstlisting}[firstnumber=7]
from pulser.devices import Chadoq2
seq = Sequence(reg, Chadoq2)
\end{lstlisting}

\subsubsection{Channel declaration}

Each device has a set of available channels, which can be consulted through \texttt{seq.available\_channels}. Since we need local addressing on both the \texttt{ground-rydberg} basis (for the CZ gate) and the \texttt{digital} basis (for the single-qubit gates), we will declare the \lstinline{'rydberg_local'} and \lstinline{'raman_local'} channels, which we will name \lstinline{'rydberg'} and \lstinline{'digital'}, respectively.

\begin{lstlisting}[firstnumber=9]
seq.declare_channel('digital', 'raman_local')
seq.declare_channel('rydberg', 'rydberg_local',
    initial_target='c')
\end{lstlisting}

On real devices, each channel can only be declared once and will no longer appear in the list of available channels after declaration, appearing instead in \texttt{seq.declared\_channels}.

\subsubsection{Making the pulses}
The different channels will be populated by \textit{pulses}, which are specific objects that need to be created. A pulse is made up of an amplitude (i.e. Rabi frequency) waveform, a detuning waveform and a fixed phase, although special convenience methods exist when one or both waveforms are constant. In our example, all pulses will be resonant, so we will use \texttt{Pulse.ConstantDetuning()} to conveniently set $\delta=0$ without defining its waveform. 

For the amplitude waveform, eq. (\ref{rot_angle}) tells us that what defines the angle of rotation is simply the integral of the waveform. This ideal scenario does not distinguish between a rectangular waveform or something more complex, as long as its integral is the same. In practice, a properly-shaped waveform can mitigate unwanted modifications to the modulated signal, stemming from spectral leakage or noise sensitivity. Here, we will use the \textit{Blackman} waveform, which is a tapering function designed to have close to minimum spectral leakage.

Starting with the single-qubit gates, all of those involved have $\theta=\pi/2$, so we declare a Blackman waveform with an area of $\pi/2$ and duration of $200$ ns, making up a relatively short pulse without exceeding the amplitude range allowed by the channel where it will be executed. 

\begin{lstlisting}[firstnumber=12]
from pulser.waveforms import BlackmanWaveform
half_pi_wf = BlackmanWaveform(200, np.pi/2)
\end{lstlisting}

The rotation axis is defined from eq. (\ref{rot_axis}), from which we gather that a rotation around the $y$-axis corresponds to a phase $\varphi = -\pi/2$. Inversely, since $R_y^\dag(\pi/2) = R_{-y}(\pi/2)$, we will have $\varphi=\pi/2$ defining a rotation around $-y$. The corresponding pulse declarations are:

\begin{lstlisting}[firstnumber=14]
ry_pulse = Pulse.ConstantDetuning(
    amplitude=half_pi_wf,
    detuning=0,
    phase=-np.pi/2,
)
ry_dag_pulse = Pulse.ConstantDetuning(
    amplitude=half_pi_wf,
    detuning=0,
    phase=np.pi/2,
)
\end{lstlisting}

Moving on to the CZ gate, which follows the protocol of Figure \ref{fig:cz_protocol}, we will need a $\pi$-pulse and a $2\pi$-pulse, which we can shape with a Blackman waveform as well. However, because we are now dealing with Rydberg interactions, we have to make sure that the Rydberg blockade radius resulting from the chosen waveform is well above the spacing between the atoms. This concern does not extend to all the pulses assigned to the \texttt{Rydberg} channel, because a Rydberg blockade is only meaningful if multiple atoms are being excited to the Rydberg state simultaneously or if an atom is being excited after its neighbors. Thus, in this case, only the $2\pi$-pulse will determine a relevant Rydberg blockade radius, which means the $\pi$-pulses surrounding it can be kept short:

\begin{lstlisting}[firstnumber=24]
pi_wf = BlackmanWaveform(200, np.pi)
pi_pulse = Pulse.ConstantDetuning(pi_wf, 0, 0)
\end{lstlisting}

To figure out what is the Rabi frequency upper bound that corresponds to a certain Rydberg blockade radius, as dictated by eq. (\ref{eq:blockade_radius}), we use the \texttt{rabi\_from\_blockade()} method of our chosen device. We set a conservative Rydberg blockade radius of $8 \mu$m, double the distance between the atoms. The corresponding maximum Rabi frequency value is then

\begin{lstlisting}[firstnumber=26]
max_val = Chadoq2.rabi_from_blockade(8)
# Result: max_val = 19.10672378540039 rad/us
\end{lstlisting}
Conversely, it is also possible to figure out what the Rydberg blockade radius is for a given Rabi frequency value through the \texttt{rydberg\_blockade\_radius()} method.

\begin{figure}[h]
    \centering
    \includegraphics[width=\linewidth]{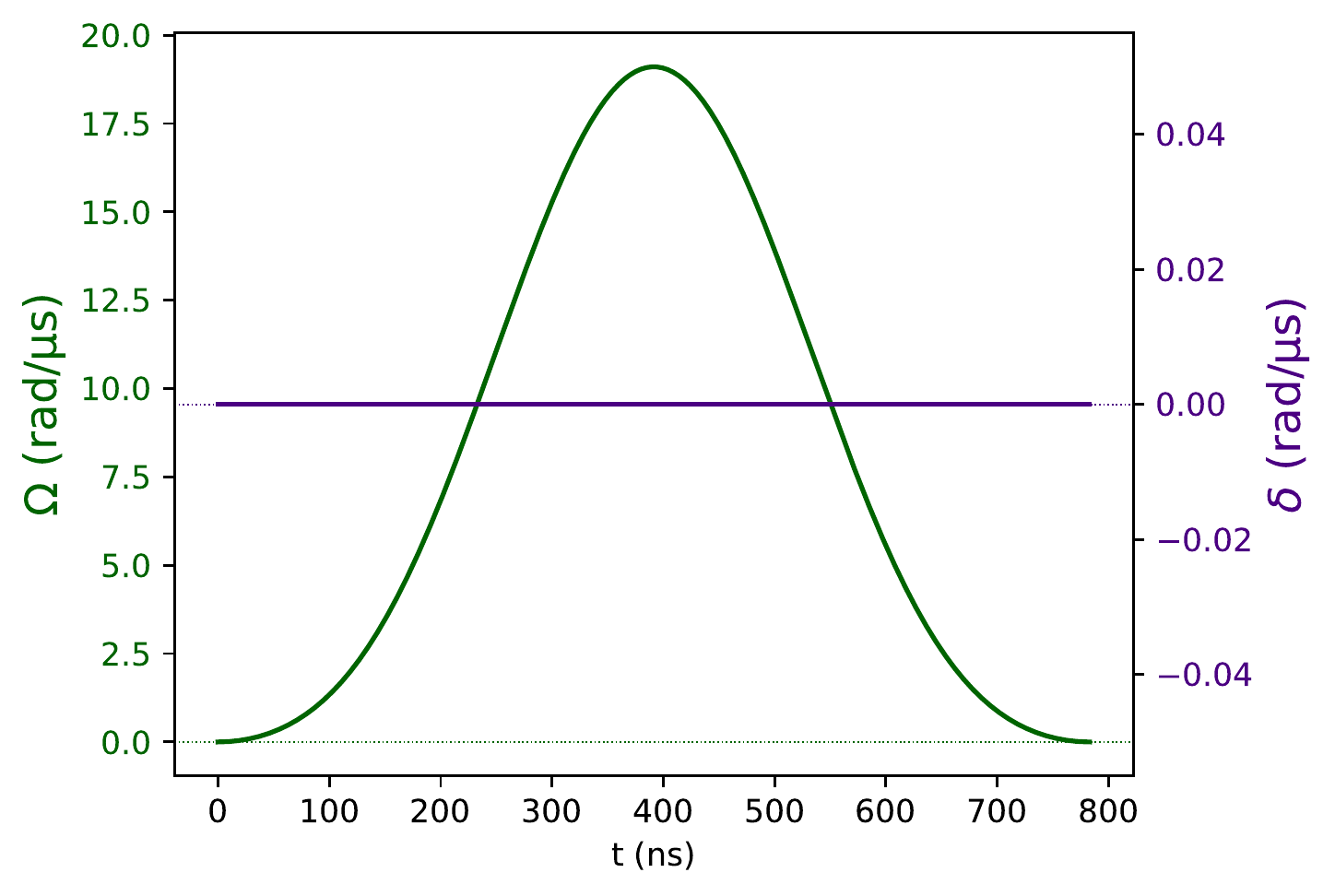}
    \caption{The outcome of calling \lstinline{two_pi_pulse.draw()}. Both the amplitude and detuning waveforms are displayed.}
    \label{fig:two_pi_draw}
\end{figure}

With an upper bound on the amplitude, we can create the appropriate Blackman waveform and corresponding pulse using the following formulation (which leverages the fact that there is a simple relationship between the duration, area and maximum value of a Blackman waveform):
\begin{lstlisting}[firstnumber=28]
two_pi_wf = BlackmanWaveform.from_max_val(
    max_val=max_val,
    area=2*np.pi,
)
two_pi_pulse = Pulse.ConstantDetuning(
    amplitude=two_pi_wf,
    detuning=0,
    phase=0,
)
\end{lstlisting}

All the pulse objects necessary to program the circuit of Figure \ref{fig:bell_circuit}(b) have been created. As a remark, note that every \texttt{Pulse} and \texttt{Waveform} can be plotted by calling its \texttt{draw()} method, which can be useful to visualize the shape of the waveforms, as well as inspect the range of values they span. As an example, Figure \ref{fig:two_pi_draw} displays the plot of \lstinline{two_pi_pulse}, showing both the amplitude and detuning waveforms in the same picture.

\subsubsection{Composing the Sequence}

As previously mentioned, the workings of neutral-atom devices ultimately influence the sequences designed in Pulser. In many aspects --- like the way single-qubit gates are realized as pulses --- neutral-atom devices are very similar to other popular implementations. However, when compared to superconducting qubit architectures for instance, there are some marked differences, the most striking being in the connection between channels and qubits. 

In superconducting qubit devices, each qubit has multiple dedicated channels, each serving a specific function, like that of manipulating or measuring the state of the qubit. This means that if one is to apply a pulse on a given qubit, that pulse will be allocated to the channel exclusively serving that qubit. 

On the other hand, a channel in a neutral-atom QPU can serve all the qubits in the register, be it simultaneously or sequentially. In practice, this means that a channel needs to know on which qubit (or qubits) to apply the pulses to, and while there are some channels --- specifically, those of the \texttt{Global} type --- whose targets are fixed to be the entire register, the channels of type \texttt{Local} have the ability to change their target throughout the sequence. When the state of a single-qubit is changed, a \texttt{Local} channel targets it before applying the pulse, after which it can move on to target a different qubit onto which the next pulse will be applied. In this way, a \texttt{Local} channel can individually manipulate the state of all the qubits in a register, though it will do it sequentially rather than in parallel.

Thus, every channel needs to start with a target. For \texttt{Global} channels, all the atoms are the target by definition, but for \texttt{Local} channels, their target has to be explicitly defined. This initial target can be set at channel declaration (see how \lstinline{'rydberg'} was set to target the \lstinline{'c'} atom), or it can be done by using the standard \lstinline{target} method of \lstinline{seq} --- the \texttt{Sequence} instance previously created.

\begin{lstlisting}[firstnumber=37]
seq.target('c', 'digital')
\end{lstlisting}

Now both channels have an initial target, so we can compose the sequence by allocating the \texttt{Pulse}s to the declared \texttt{Channel}s.

We start with the single-qubit gates, $R_y(\pi/2)$ and $R_y^\dag(\pi/2)$. Recall that these transitions are addressed by the \texttt{Raman}-type channel, which we called \lstinline{'digital'}. Thus, the first two gates of circuit \ref{fig:bell_circuit} are added to the same channel as follows:

\begin{lstlisting}[firstnumber=38]
seq.add(ry_pulse, 'digital')
seq.target('t', 'digital')
seq.add(ry_dag_pulse, 'digital')
\end{lstlisting}
Notice how the target is changed to \lstinline{'t'} in between additions, so that the $R_y^\dag(\pi/2)$ gate acts on the correct qubit. 

Next, we implement the CZ gate by replicating the protocol of Figure \ref{fig:cz_protocol}. This will involve the other declared channel, so we must clarify how pulses in different channels are aligned in time. By default, when adding a pulse to a channel (which is targeting one or more qubits), it will be added as soon as all targeted qubits are free. This means that, if existing pulses on other channels are acting on one of the targeted qubits, the channel will be idle until they are done. On the other hand, if there is no overlap between the different channels' targets, the pulses will be played in parallel; this is called the \lstinline{'min-delay'} protocol. The other addition protocols are the \lstinline{'wait-for-all'}, in which the pulse will only play after all previously added pulses are finished, and the \lstinline{'no-delay'}, where a delay is never added. 

There is an additional command, called \texttt{align()}, which introduces delays that align a subset of channels with the one that finished the latest, such that the next pulse added to any of them will start right after the latest channel has finished. When all the declared channels are aligned, the outcome is identical to that of using the \lstinline{'wait-for-all'} protocol; otherwise, when only some of the channels are aligned, it corresponds to a ``wait-for-some'' protocol.

As illustrated in Fig.\,\ref{fig:cz_protocol}, the CZ implementation starts with a $\pi$-pulse on channel \lstinline{'rydberg'}, which is targeted to the \lstinline{'c'} qubit. The channel starts empty, but the pulse will not be added at $t=0$ by default because it would overlap with the first pulse of channel \lstinline{'digital'}, which is also targeting \lstinline{'c'}. Instead, it would wait for this first pulse to end and then start playing in parallel with the second pulse in the \lstinline{'digital'} channel. Although this would be a valid implementation, we will instead use the \lstinline{align} command to make the pulse start only after the \lstinline{'digital'} channel is done with both pulses (the \lstinline{'wait-for-all'} would work too because we only declared two channels, but this way can be generalized to the case were other channels are declared). This will avoid dead-times in the CZ implementation, thus keeping the atoms in the Rydberg state no longer than necessary (which is desirable, since the atoms are more sensitive to noise when in the Rydberg state). 

Thus, the CZ is implemented through
\begin{lstlisting}[firstnumber=41]
seq.align('digital', 'rydberg')
seq.add(pi_pulse, 'rydberg')
seq.target('t', 'rydberg')
seq.add(two_pi_pulse, 'rydberg')
seq.target('c', 'rydberg')
seq.add(pi_pulse, 'rydberg')
\end{lstlisting}

Finally, there is still a single-qubit gate on the \lstinline{'t'} qubit that must be applied after the CZ is done (recall that channel \lstinline{'digital'} was left targeting \lstinline{'t'}),

\begin{lstlisting}[firstnumber=47]
seq.align('digital', 'rydberg')
seq.add(ry_pulse, 'digital')
\end{lstlisting}

\begin{figure*}
\centering
\includegraphics[width=\linewidth]{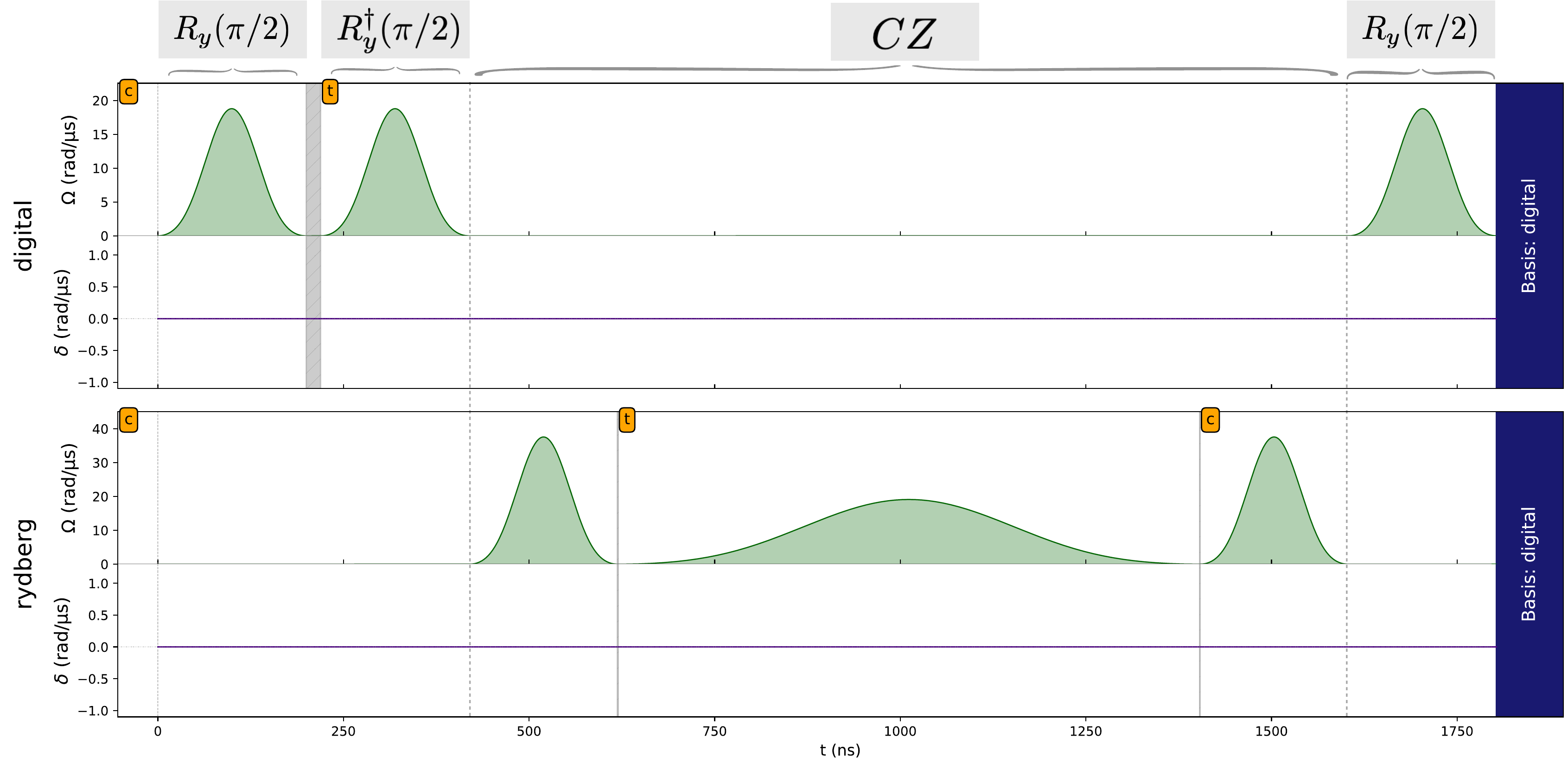}
\caption{The final Bell state creation sequence, displayed by calling \lstinline{seq.draw()}. The current target of each channel is displayed in the orange boxes and is part of the original output, but the gates corresponding to each part of the circuit were annotated on top of it. The blue boxes on the right specify the basis in which the system is measured at the end of the sequence.}
\label{fig:cz_seq}
\end{figure*}

\subsubsection{Measurement}

At last, the measurement signals the end of a sequence, so after it no more changes are possible. We can measure a sequence by calling:
\begin{lstlisting}[firstnumber=49]
seq.measure(basis='digital')
\end{lstlisting}
When measuring, one has to select the desired measurement basis\footnote{Here, we choose the \lstinline{'digital'} basis, not to be confused with the channel we named \lstinline{'digital'} for addressing this transition.}. The available options depend on the device and can be consulted by calling the chosen device's \texttt{supported\_bases} attribute.

\subsubsection{Visualization}

Throughout its creation process, the sequence can be visualized through the \texttt{Sequence.draw()} method.
As an example, the final sequence is shown in Figure \ref{fig:cz_seq}. Note that the phase of each pulse is not represented, only the amplitude and detuning modulations of each channel's output --- the latter being always zero in this example. To inspect the phase of each pulse, one can print out the sequence instead, through
\begin{lstlisting}[firstnumber=50]
print(seq)
\end{lstlisting}
This will list all the contents of each channel, alongside its start and finish times.

\subsubsection{Simulation}
The primary goal of Pulser is to serve as a front-end framework for designing pulse sequences for neutral-atom quantum processors. However, it is also possible to emulate the output of a sequence locally on a classical device through the \texttt{Simulation} class. Pulser's emulator relies on the open-source package QuTiP. Incorporating an emulator in Pulser is valuable for several reasons. First, it enables users to carefully design their pulse sequences and understand their effects before sending the instructions to a real device. It also enables them to gain intuition with simple setups, and explore new possible applications on limited-size systems. Coming back to our example, we use the emulator on our sequence by calling:
\begin{lstlisting}[firstnumber=51]
from pulser.simulation import Simulation
sim = Simulation(seq)
res = sim.run()
\end{lstlisting}
Every time the \texttt{run()} method is called, a \texttt{SimulationResults} object instance is returned, which holds the results of the simulation.

The simulation performed is a time evolution of the system at specific times, such that \lstinline{res} holds the state of the system at each instant. All the states can be accessed through the \texttt{states} attribute, the final entry being the state of the system at the end of the sequence (before measurement). 

In this case, since the three energy levels of the two atoms were involved, the final state will be a nine-entry vector. However, the population of the Rydberg levels should be negligible, so it is possible to reduce the final state to the \lstinline{'digital'} basis of interest. We do so by calling
\begin{lstlisting}[firstnumber=54]
res.get_final_state(reduce_to_basis='digital')
\end{lstlisting}
which returns a \texttt{qutip.Qobj} instance holding the final state in the \lstinline{'digital'} basis. For this example, the outcome is
$$
\begin{pmatrix} 
    0.707+i0.006 \\
    -3.88 \times 10^{-05}+i0.006 \\
    -2.99\times 10^{-05}      \\
     0.707       
\end{pmatrix},
$$
which is very close to what we expect for $\ket{\Phi^+}$. 

Another way to probe the results is to emulate a real experiment, in which case the only information about the final state of the system is obtained in the measurement. For a given number \texttt{N\_samples} of repetitions of the experiment, we call
\begin{lstlisting}[firstnumber=55]
res.sample_final_state(N_samples=1e4)
\end{lstlisting}
Due to finite sampling errors, the outcome will vary slightly for each call. An example of an outcome is \lstinline[] ${'00': 5005, '11': 4995}$.

\section{Pulser in action : exploration of usecases with the emulator}
\label{section:usecases}

In this section, we reproduce some results obtained experimentally using Pulser's built-in emulator. These examples, among several others, can be  accessed online at Pulser's documentation site\footnote{https://pulser.readthedocs.io}, which includes the code for the optimization of parameters and the generation of the figures.

\begin{figure*}[t!]
\centering
\includegraphics[width=\linewidth]{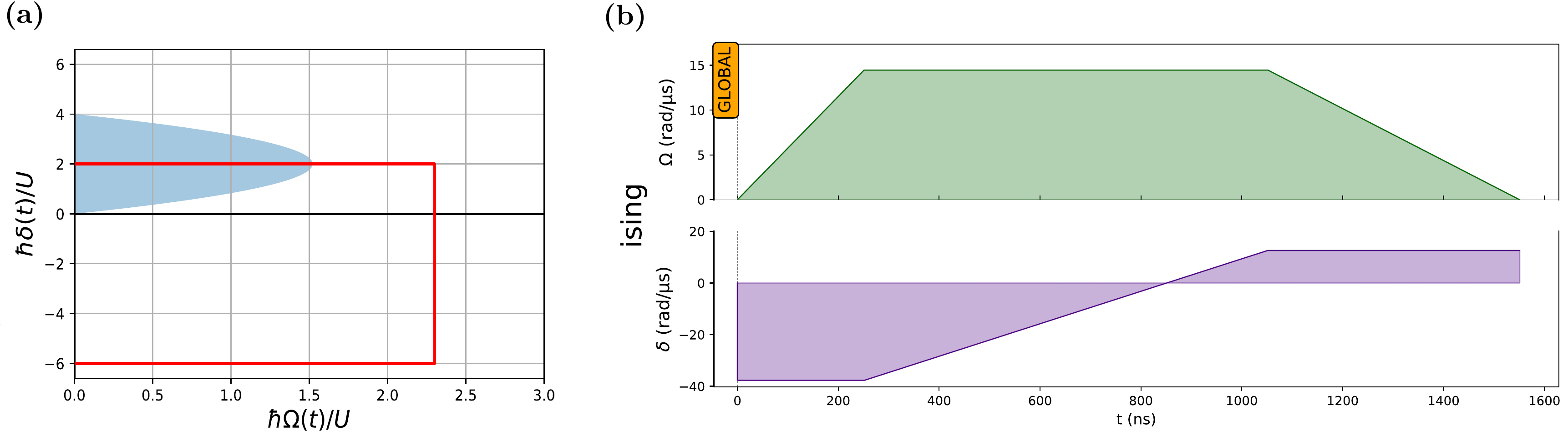}
\caption{(a) Schematic phase diagram of a two-dimensional quantum Ising system, with respect to the ratio of applied Rabi frequency $\hbar \Omega/U$ and detuning $\hbar \delta/U$. The blue region indicates the antiferromagnetic (AFM) phase, and outside of this region a paramagnetic (PM) phase is found. (b) The pulse sequence for the adiabatic preparation of the antiferromagnetic state in the Ising model, displayed by calling \lstinline{seq.draw()}.}
\label{AFM:phase_diagram_and_seq}
\end{figure*}

\subsection{Preparing an antiferromagnetic state and observing the produced correlations}
\label{adiabatic_preparation}

First we illustrate how to build a sequence for adiabatically preparing an antiferromagnetic state from the Ising-like model of eq.\eqref{eq:ising_ham}, as well as studying its correlations. These types of states have been studied in neutral-atom setups up to hundreds of qubits, showing remarkable precision and allowing measurements that are beyond the best available classical simulation algorithms \cite{Bernien17,Lienhard18, scholl2020programmable, ebadi2020quantum}.

The atoms will be placed in a square array of interatomic distance $R$. We will shine a laser globally with a pulse sequence representing a path in the phase diagram of the quantum Ising model, that starts in the paramagnetic phase and ends in the antiferromagnetic dome, as shown in Fig. \ref{AFM:phase_diagram_and_seq}(a). We shall evaluate the quality of the resulting state by calculating the spin-spin correlation function, defined as:
\begin{equation}\label{eq:corr_fct}
g^{(2)}(k,l) = \frac{1}{N_{k,l}}\sum_{\{i,j\}} \Big(\langle \hat n_i \hat n_j \rangle - \langle \hat n_i \rangle \langle \hat n_j \rangle \Big).
\end{equation}
The function $g^{(2)}$ measures the correlation between local operators $\hat n_i, \hat n_j$ at different positions, averaged over all pairs of atoms $i,j$ separated by a vector $(kR,lR)$ ($N_{k,l}$ is the total number of such pairs). The pulses will be constructed by indicating the values for the maximum Rabi frequency, and the initial and final detuning (as reported in \cite{lienhard2018observing}). 
\begin{lstlisting}[firstnumber=1]
# Load basic packages and classes
import numpy as np
import pulser
from pulser import Register, Pulse, Sequence
from pulser.devices import Chadoq2
from pulser.waveforms import RampWaveform
from pulser.simulation import Simulation
# Define setup parameters (in rad/us and ns)
Omega_max = 2.3 * 2*np.pi
U = Omega_max / 2.3
delta_0 = -6 * U
delta_f = 2 * U
t_rise = 250
t_fall = 500
t_sweep = (delta_f-delta_0)/(2*np.pi*10)*1000
\end{lstlisting}
Since we want to use the blockade effect for nearest-neighbors only, we tune $\Omega_\text{max}$ to be of the same order as the van der Waals interaction magnitude for nearby atoms, $\Omega_{\text{max}} \sim U$. This will fix an interatomic distance $R_{\text{interatomic}}$, which we use to define our register:

\begin{lstlisting}[firstnumber=16]
R_interatomic = Chadoq2.rydberg_blockade_radius(U)
reg = Register.square(3, R_interatomic, prefix='q')
\end{lstlisting}

\subsubsection{Creating the sequence}

We compose our pulse using the \texttt{Pulse} and \texttt{Sequence} classes:

\begin{lstlisting}[firstnumber=18]
# Define each pulse
rise = Pulse.ConstantDetuning(
        RampWaveform(t_rise, 0., Omega_max),
        delta_0,
        0.)
sweep = Pulse.ConstantAmplitude(
        Omega_max,
        RampWaveform(t_sweep, delta_0, delta_f),
        0.)
fall = Pulse.ConstantDetuning(
        RampWaveform(t_fall, Omega_max, 0.),
        delta_f,
        0.)
# Add to sequence
seq = Sequence(reg, Chadoq2)
seq.declare_channel('ising', 'rydberg_global')
seq.add(rise, 'ising')
seq.add(sweep, 'ising')
seq.add(fall, 'ising')
\end{lstlisting}
The sequence is readily plotted with the \texttt{draw()} method, as illustrated in Figure \ref{AFM:phase_diagram_and_seq}(b).

\subsubsection{Simulation: Spin-Spin Correlation Function}

\begin{figure}[h]\label{fig:AFM_hist}
\centering
\includegraphics[width=\linewidth]{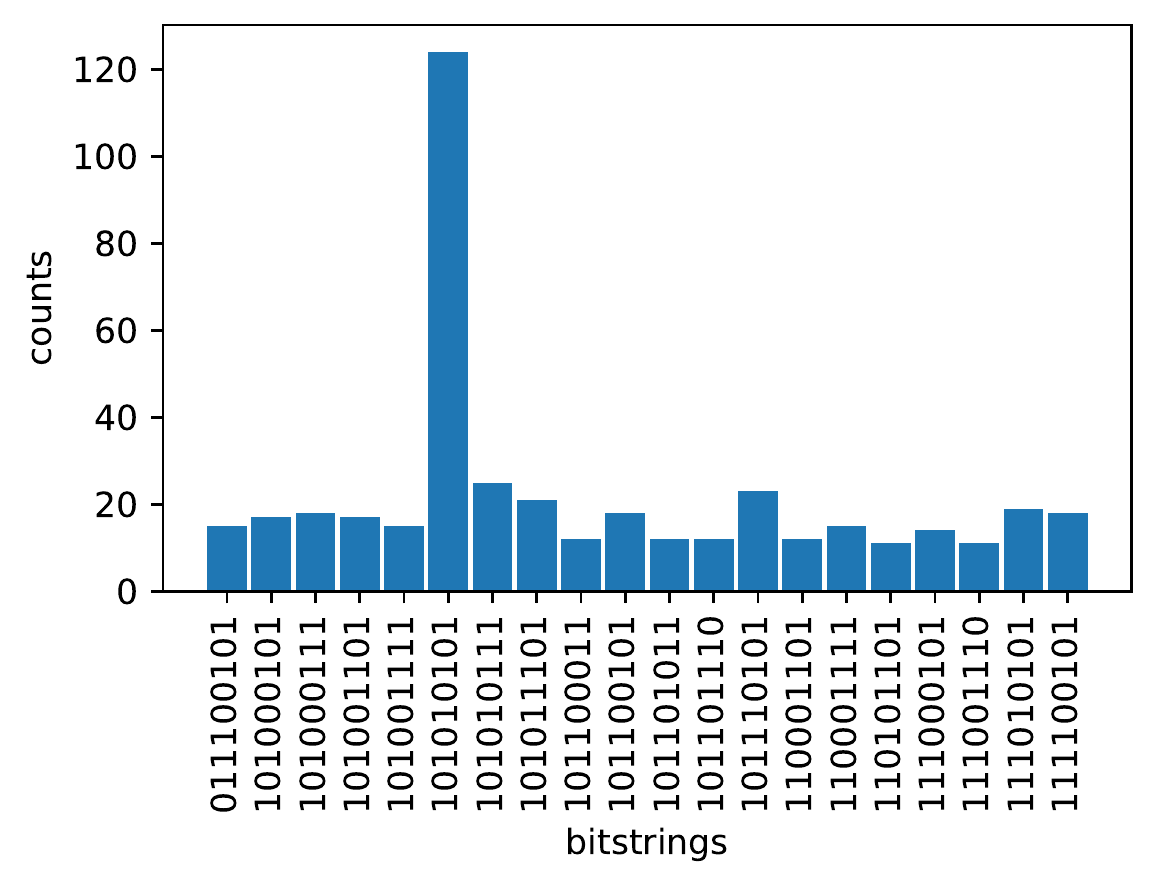}
\caption{Most frequent results from the sampling of the final state.}
\label{histogram_results}
\end{figure}

We run a simulation of the sequence with the \texttt{run()} method of the \texttt{Simulation} class. For simple pulse sequences like the one we are using here, we can use a reduced sampling rate for faster emulation. A progress estimate can also be included if desired, with the parameter \lstinline{progress_bar=True}. Once the final state is obtained, we may sample from it using the \texttt{sample\_final\_state()} method. Since there was no measurement basis (\lstinline{'ground-rydberg'} or \lstinline{'digital'}) specified during the composition of the sequence, we include it as an argument with \texttt{meas\_basis}:

\begin{lstlisting}[language=python, firstnumber=37]
simul = Simulation(seq, sampling_rate=0.02)
results = simul.run(progress_bar=True)
count = results.sample_final_state(
                   meas_basis='ground-rydberg')
most_freq = {k:v for k,v in count.items() if v>10}
\end{lstlisting}

From the output of the simulation, we can plot the histogram of the sampled results, shown in Figure \ref{histogram_results}. Notice the peak corresponding to the antiferromagnetic state.\\

\begin{figure}[h]
\centering
\includegraphics[width=\linewidth]{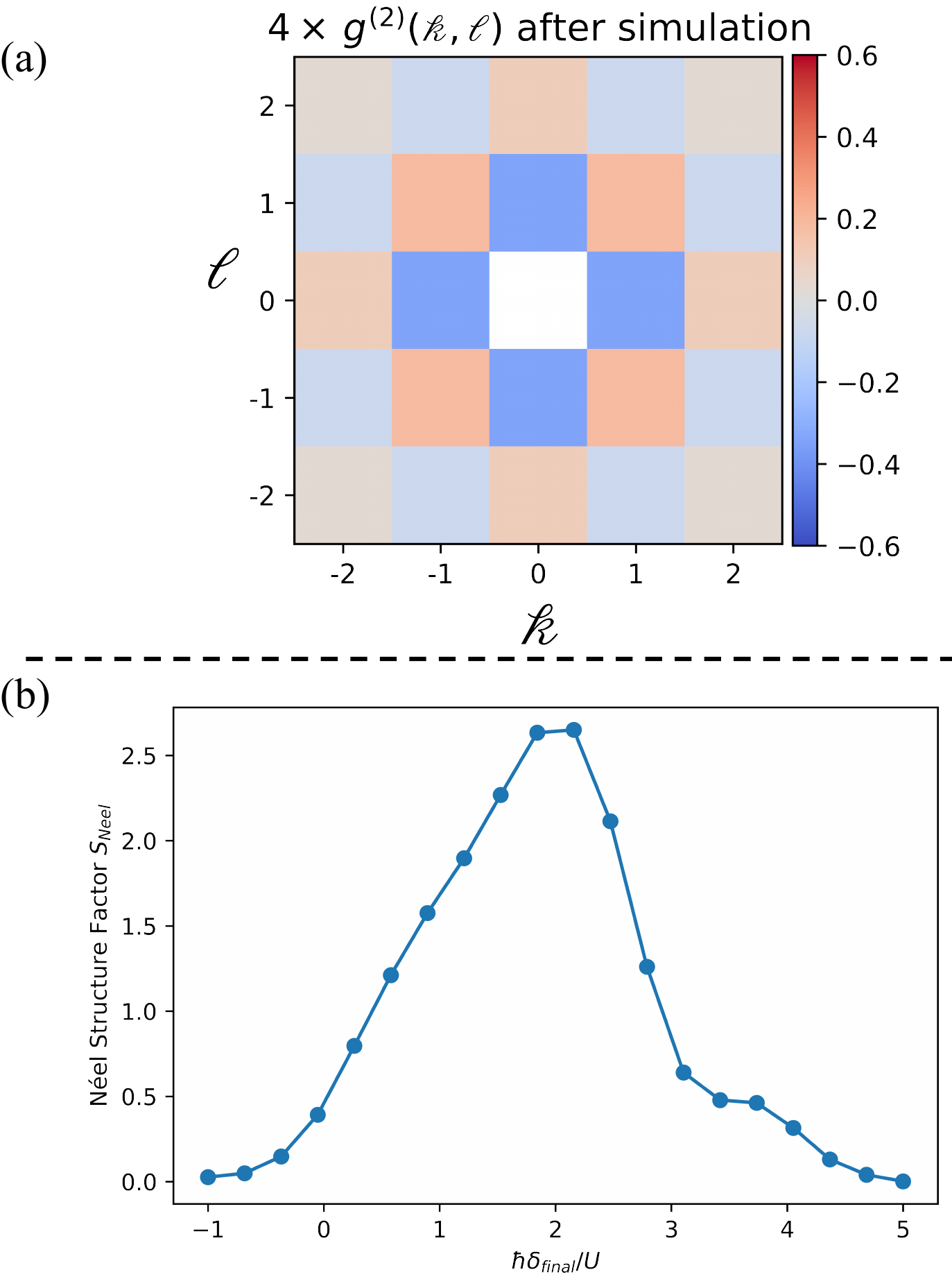}
\caption{(a) Simulation results for the $g^{(2)}$ correlation function, eq.\eqref{eq:corr_fct}. (b) Sweep results for the $S_{\text{Néel}}$ score function, eq.\eqref{eq:neel}.}
\label{fig:AFM}
\end{figure}

From \lstinline{results}, one can access the state of the system throughout the simulation (as a list of \texttt{qutip.Qobj}s) by calling \lstinline{results.states}. This can be useful for post-processing the simulation results and computing observables. Alternatively, one can also call \lstinline{results.expect()}  to get the exact expectation value of a list of observables. For example, \lstinline{results.expect( [qutip.basis(9, 0).proj()])} will return the probability of measuring the all-excited state $|r\rangle^{\otimes 9}$ for every state in \lstinline{results}.

We show in Figure \ref{fig:AFM}(a) the output of the simulation for the correlation function $g^{(2)}$, eq. \eqref{eq:corr_fct}. Note that the correlation function is expected to decay exponentially (modulo finite-size effects), which is best observed at larger system sizes~\cite{scholl2020programmable}.

Changing the endpoint $\delta_f$ of the pulse's path results in different qualities of the final state. In order to evaluate how well it represents an antiferromagnetic state, we will explore the $\Omega = 0$ line on the phase diagram, Fig. \ref{AFM:phase_diagram_and_seq}(a). Since the value of correlation function $g^{(2)}(k,l)$ (eq. \eqref{eq:corr_fct}) can be positive or negative, we compensate with alternating signs to construct a single score for the state \cite{lienhard2018observing}:
\begin{equation}
S_{\text{Néel}} = \sum_{(k,l)\neq (0,0)} (-1)^{|k| + |l|} g^{(2)}(k,l),
\label{eq:neel}
\end{equation}
which should be largest when the state is closest to an antiferromagnetic product state. By sweeping over different values of detuning $\delta_f$, we examine the region $ 0 < \hbar \delta_f/U < 4$. In a few seconds, Pulser gives the results shown in Figure \ref{fig:AFM} (b), showing where the antiferromagnetic phase is more strongly present.

\subsection{Variational algorithms with Pulser: solving a graph problem with neutral atoms}

The implementation of variational algorithms, such as the Quantum Approximation Optimization Algorithm\,\cite{Farhi14} (QAOA), is facilitated by the use of a \textit{parametrized sequence} in Pulser. We illustrate this aspect here, by solving the \emph{Maximum Independent Set} (MIS) problem on \emph{Unit Disk} graphs, which can be naturally studied on neutral-atom devices \cite{pichler2018quantum, henriet2020robustness, dalyac2020qualifying}.

\subsubsection{From a graph to an atomic register} 

An \emph{independent set} of a graph is a subset of vertices where any two elements of this subset are not connected by an edge. The MIS corresponds to the largest of such subsets. The so-called \emph{Unit Disk} (UD) version of this problem corresponds to the instances where the graph under consideration lives in 2D and displays an edge between two nodes if they are within a unit length of each other.\\

\begin{figure}[h]
\centering
\includegraphics[scale=0.6]{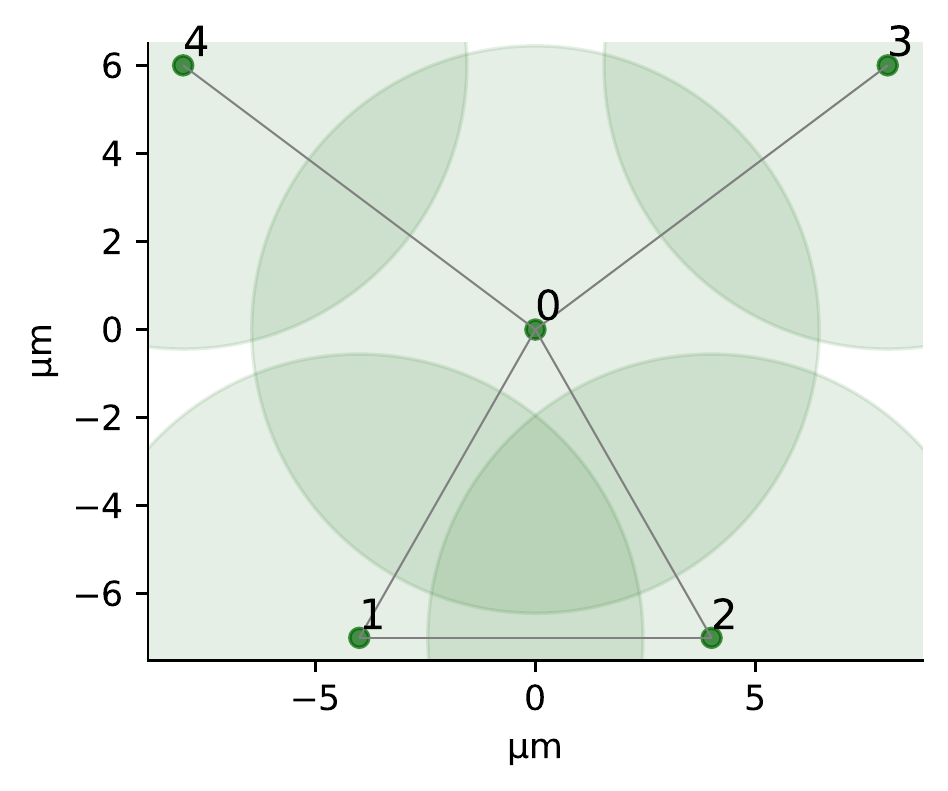}
\caption{The register for the MIS graph problem. The shaded circles show the blockade half-radius, such that the intersecting regions indicate existing links.}
\label{fig:MIS_reg}
\end{figure}

Interestingly, ensembles of interacting Rydberg atoms in 2D can be naturally represented by Unit Disk graph structures: the atoms are the nodes of the graphs and an edge in the graph corresponds to two atoms being within the blockade radius of each other. In this example, we will take $\Omega$ fixed to a frequency of 1 $\text{rad}/\mu$s, hence the blockade radius can be obtained by using \lstinline{Chadoq2.rydberg_blockade_radius(1.)}. The following code block instantiates the atomic register, and displays the graph induced by their interactions with the \lstinline{draw_graph=True} and \lstinline{draw_half_radius=True} arguments passed to the \lstinline{reg.draw()} method, as illustrated in Figure \ref{fig:MIS_reg}.

\begin{lstlisting}[firstnumber=1]
# Load basic packages and classes
import numpy as np
import pulser 
from pulser import Register, Pulse, Sequence
from pulser.devices import Chadoq2
from pulser.simulation import Simulation
# Set Register
pos = np.array([[0., 0.], [-4, -7], [4,-7], [8,6], [-8,6]])
reg = Register.from_coordinates(pos)
Rb = Chadoq2.rydberg_blockade_radius(1.)

reg.draw(
    blockade_radius=Rb,
    draw_graph=True,
    draw_half_radius=True
    )
\end{lstlisting}

The graph $G$ has two maximal independent sets: $(1,3,4)$ and $(2,3,4)$, respectively \texttt{01011} and \texttt{00111} in binary. One could try to prepare those states using an adiabatic approach such as the one illustrated in section \ref{adiabatic_preparation}. Another approach is to use QAOA, as we illustrate below.

\subsubsection{Building the parametrized sequence}

QAOA exploits both a quantum and a classical processor that exchange information within a feedback loop to solve an optimization problem. The role of the neutral-atom processor is to prepare and measure an $N$-atom parametrized wavefunction. The outcome of the measurement is then used by the classical processor in a standard classical optimization procedure, that updates the parameters for the next iteration.\\

More specifically, the preparation of the parametrized wavefunction is achieved through the successive application of two non-commutative Hamiltonians, with all atoms initially starting in the ground state $|00\dots0\rangle$ of the \lstinline{ground-rydberg} basis. The first one is realized by taking $\Omega = 1$ $\text{rad}/\mu s$, and $\delta = 0 $ $\text{rad}/\mu s$ in Eq.(\ref{eq:global_ising}). The second Hamiltonian is realized with $\Omega = \delta = 1$ $\text{rad}/\mu s$. These two Hamiltonians are applied successively, for durations $t_j$ and $s_j$, respectively. The dimension of $\bm t$ and $\bm s$, i.e. the number of layers applied to the system, is referred to as the depth of the algorithm. The outcome of the measurement is then used as the objective function in a standard classical optimization procedure, that updates the parameters for the next iteration.\\

When implementing a variational procedure, multiple sequences are created that differ only by the durations $t_j$ and $s_j$ of the pulses. This is conveniently handled in Pulser by turning a regular sequence into a \textit{parametrized} sequence. A \texttt{Sequence} is said to be parametrized once a \textit{variable} is declared and used in a sequence-building call, from which point the sequence building process can continue as usual but the sequence is no longer built on the fly. Instead, all calls to a parametrized sequence are stored as a \textit{blueprint} for generating a new \texttt{Sequence}, which may depend on the value of the declared variables. Consequently, it is not possible to progressively monitor the creation process of a sequence that is parametrized (e.g. by drawing the state of a sequence as new pulses are added). The building of the sequence itself only happens when \texttt{Sequence.build()} is called, at which point specific values for all the declared variables have to be specified as arguments.\\

The \texttt{Variable} objects are obtained by calling \texttt{Sequence.declare\_variable()} and can be used throughout the \texttt{Sequence} creation process (i) as parameters when creating new \texttt{Waveform}s and \texttt{Pulses}, or (ii) as arguments for standard sequence creation methods like \texttt{add}, \texttt{target}, \texttt{align} and \texttt{delay}. Moreover, they support basic arithmetic operations and, when they are of \texttt{size}$>1$, iteration and item accessing. \\

The notion of parametrized sequence can become very handy with real world constraints such as those we find in cloud-based platforms. First, the bandwidth allocation of a program, variational or not, is greatly reduced thanks to this factorization. A user only needs to send the parametrized sequence that describes his program once, and then each new iteration only requires the associated set of parameters. Thanks to the high flexibility of Pulser, users can have a very fine-grained control over the waveforms that define their pulses. This results in needing larger objects to define these pulses, which makes such a factorization even more useful.

In addition, one area of improvement for cold atom platforms is the large calibration times for atom register configuration changes. This means that running randomly independent sequences on the same QPU can be very inefficient. The sequences of a given user's program on the other hand share some common parameters, including the register configuration. On the QPU side, we can treat the parametrized sequences as the basic block of the QPU scheduling to greatly reduce the latencies due to the calibration processes, make sure the different user programs are run efficiently and that all the sequences in a given program are run on the same QPU.\\

The example below shows how \texttt{Sequence} can be used to create a parametrized QAOA sequence with two layers, with the variable duration of each layer stored in the \lstinline{t_list} and \lstinline{s_list} arrays. Notice that no value is given to the contents of \lstinline{t_list} and \lstinline{s_list}.

\begin{lstlisting}[firstnumber=17]
# Parametrized sequence
seq = Sequence(reg, Chadoq2)
seq.declare_channel('ch0', 'rydberg_global')

t_list = seq.declare_variable('t_list', size=2)
s_list = seq.declare_variable('s_list', size=2)

for t, s in zip(t_list, s_list): 
    pulse_1 = Pulse.ConstantPulse(1000*t, 1., 0., 0) 
    pulse_2 = Pulse.ConstantPulse(1000*s, 1., 1., 0)
    
    seq.add(pulse_1, 'ch0')
    seq.add(pulse_2, 'ch0')
\end{lstlisting}

The sequence \lstinline{seq} above will only be built once the user provides specific values for \lstinline{t_list} and \lstinline{s_list} while calling the \texttt{Sequence.build()} method. It thus enables to build a variety of sequences that share the same structure:

\begin{lstlisting}[numbers=none]
# Build sequences with specific values
my_seq_1 = seq.build(t_list=[2,4], s_list=[3,6])
my_seq_2 = seq.build(t_list=[1,3], s_list=[2,5])
\end{lstlisting}

\subsubsection{Classical optimization and QAOA results}

The parametrized sequence above can then be used in conjunction with a  classical optimizer in order to determine a set of good values for \lstinline{t_list} and \lstinline{s_list}. When running the full QAOA procedure in closed loop, the optimizer is responsible for iteratively selecting the next set of parameters to be tested. The parametrized sequence can then be updated in Pulser and the new program sent externally to the quantum hardware. 

\begin{figure}[h]
\centering
\includegraphics[width=0.9\linewidth]{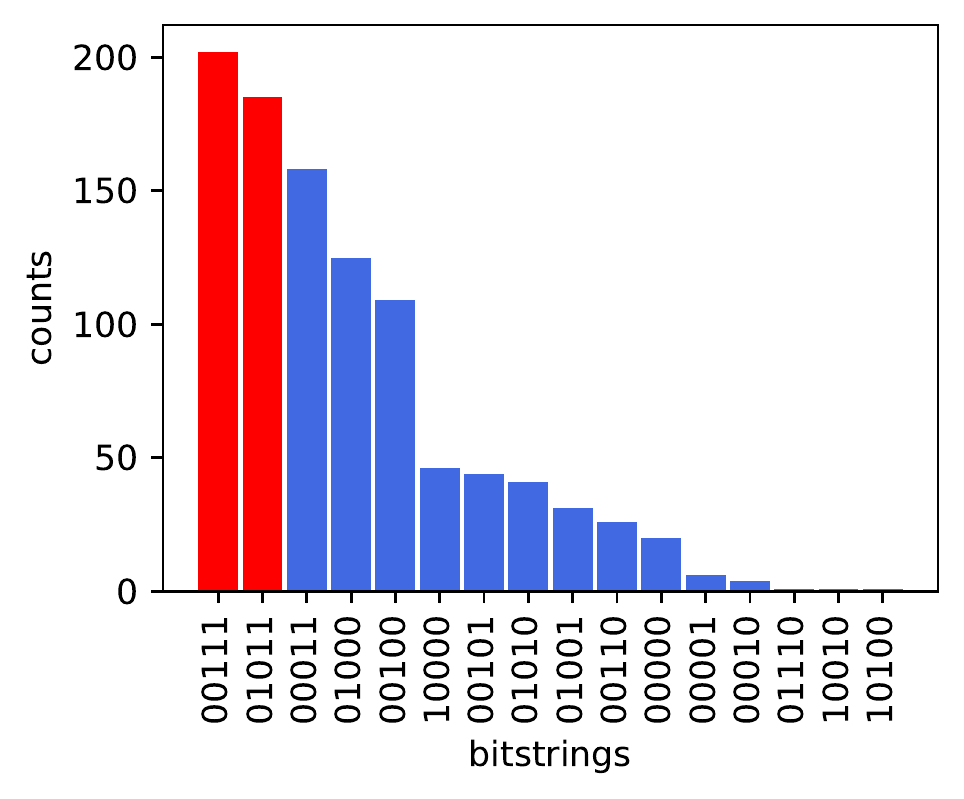}
\caption{Histogram of most frequent bitstrings after the optimization loop. The MIS graphs (in red) are the ones that are observed most frequently.}
\label{fig:MIS_optim}
\end{figure}

The procedure can also be emulated locally: to try out the algorithm, we applied the non-gradient method Nelder-Mead for a few dozen function evaluations, initializing the parameters in a convenient point (the implementation can be found at Pulser's online documentation). 
This is already sufficient in order to find some acceptable parameters $\bm t$ and $\bm s$. The performance of QAOA can be tested by sampling from the final state $|\psi(t_f)\rangle$ which returns both MISs of the graph with high probability. We show in Figure \ref{fig:MIS_optim} the histogram of the recorded bitstrings.

\section{Conclusion}

With recent advances in the field, neutral-atom processors now provide unique opportunities for the exploration of uncharted territories in many-body quantum physics with hundreds of interacting quantum units\,\cite{ebadi2020quantum,scholl2020programmable,Bluvstein21}. By using distinct electronic transitions for encoding the quantum information, one can implement various Hamiltonians depending on the task at hand. Beyond analog computing, recent developments in qubit addressing have opened the door to high-fidelity quantum gates\,\cite{levine_high-fidelity_2018,Levine19}. In addition to the flexibility in terms of information encoding and processing, the geometry of the quantum register itself is also highly programmable\,\cite{barredo_synthetic_2018,Schymik20}. This holds great promises for the implementation of graph-related algorithms\,\cite{Pichler18,zhou20,henriet2020robustness,dalyac2020qualifying,Henry21}, the exploration of lattice models in condensed-matter physics\,\cite{ebadi2020quantum,scholl2020programmable,leseleuc_observation_2019}, or the reduction of the gate-count for quantum circuits\,\cite{Henriet2020quantum}. 

In this paper, we introduced Pulser, an open-source Python library allowing users to pilot such powerful devices at the pulse level. This tool will enable the exploration of new avenues in physics and quantum information science requiring an advanced control of the system. After their creation in Pulser, pulse sequences can be translated and read by an arbitrary waveform generator for implementation on a quantum processor. The architecture of the software makes it possible to use as an interface for any neutral-atom QPU as the specifics of the back-end can be encapsulated by creating new \texttt{Device} objects. Pulser currently enables the control of laser pulses during the quantum information processing phase, but it could be extended in the future to include other controls either in the processing phase, during the assembly of the register or the measurement phase. Being a low-level tool, Pulser could be interfaced with higher-level application-oriented libraries and frameworks for digital quantum information processing, such as Cirq\,\cite{cirq_developers_2021_4586899}, Qiskit\,\cite{Qiskit}, Atos myQLM, Pennylane\,\cite{Pennylane} or tket\,\cite{tket}, which abstract away many hardware features. 

The primary purpose of Pulser's built-in emulator is to reproduce faithfully what the output of a real experiment would look like, in order to facilitate the design of hardware runs. In that spirit, some developments are currently underway for the inclusion of realistic error models\,\cite{sylvain18}. Some work could also be done to extend the simulations capabilities in terms of the number of qubits that can be efficiently simulated, including the development of tensor-network based methods and GPU acceleration. 

\section*{Acknowledgments}
We thank Lucas Beguin, Julien Brémont, Antoine Browaeys, Mauro d'Arcangelo, Eric Giguere, Christophe Jurczak, Thierry Lahaye, Boxi Li, Georges-Olivier Reymond, Pascal Scholl, Adrien Signoles, and William J. Zeng for useful discussions.

\bibliographystyle{unsrtnat}
\bibliography{refs}

\begin{thebibliography}{60}
\providecommand{\natexlab}[1]{#1}
\providecommand{\url}[1]{\texttt{#1}}
\expandafter\ifx\csname urlstyle\endcsname\relax
  \providecommand{\doi}[1]{doi: #1}\else
  \providecommand{\doi}{doi: \begingroup \urlstyle{rm}\Url}\fi

\bibitem[{Saffman} et~al.(2010){Saffman}, {Walker}, and {M{\o}lmer}]{Saffman10}
M.~{Saffman}, T.~G. {Walker}, and K.~{M{\o}lmer}.
\newblock {Quantum information with Rydberg atoms}.
\newblock \emph{Reviews of Modern Physics}, 82\penalty0 (3):\penalty0
  2313--2363, Jul 2010.
\newblock \doi{10.1103/RevModPhys.82.2313}.

\bibitem[{Saffman}(2016)]{Saffman2016}
M.~{Saffman}.
\newblock {Quantum computing with atomic qubits and {R}ydberg interactions:
  progress and challenges}.
\newblock \emph{Journal of Physics B Atomic Molecular Physics}, 49\penalty0
  (20):\penalty0 202001, October 2016.
\newblock \doi{10.1088/0953-4075/49/20/202001}.

\bibitem[Barredo et~al.(2016)Barredo, L{\'e}s{\'e}leuc, Lienhard, Lahaye, and
  Browaeys]{barredo_atom-by-atom_2016}
Daniel Barredo, Sylvain~de L{\'e}s{\'e}leuc, Vincent Lienhard, Thierry Lahaye,
  and Antoine Browaeys.
\newblock An atom-by-atom assembler of defect-free arbitrary two-dimensional
  atomic arrays.
\newblock \emph{Science}, 354\penalty0 (6315):\penalty0 1021--1023, November
  2016.
\newblock ISSN 0036-8075, 1095-9203.
\newblock \doi{10.1126/science.aah3778}.
\newblock URL \url{https://science.sciencemag.org/content/354/6315/1021}.

\bibitem[Endres et~al.(2016)Endres, Bernien, Keesling, Levine, Anschuetz,
  Krajenbrink, Senko, Vuletic, Greiner, and Lukin]{endres_atom-by-atom_2016}
Manuel Endres, Hannes Bernien, Alexander Keesling, Harry Levine, Eric~R.
  Anschuetz, Alexandre Krajenbrink, Crystal Senko, Vladan Vuletic, Markus
  Greiner, and Mikhail~D. Lukin.
\newblock Atom-by-atom assembly of defect-free one-dimensional cold atom
  arrays.
\newblock \emph{Science}, 354\penalty0 (6315):\penalty0 1024--1027, November
  2016.
\newblock ISSN 0036-8075, 1095-9203.
\newblock \doi{10.1126/science.aah3752}.
\newblock URL \url{https://science.sciencemag.org/content/354/6315/1024}.

\bibitem[Barredo et~al.(2018)Barredo, Lienhard, L{\'e}s{\'e}leuc, Lahaye, and
  Browaeys]{barredo_synthetic_2018}
Daniel Barredo, Vincent Lienhard, Sylvain~de L{\'e}s{\'e}leuc, Thierry Lahaye,
  and Antoine Browaeys.
\newblock Synthetic three-dimensional atomic structures assembled atom by atom.
\newblock \emph{Nature}, 561\penalty0 (7721):\penalty0 79--82, September 2018.
\newblock ISSN 1476-4687.
\newblock \doi{10.1038/s41586-018-0450-2}.
\newblock URL \url{https://www.nature.com/articles/s41586-018-0450-2}.

\bibitem[Browaeys and Lahaye(2020)]{Browaeys20}
Antoine Browaeys and Thierry Lahaye.
\newblock Many-body physics with individually controlled {R}ydberg atoms.
\newblock \emph{Nature Physics}, 16\penalty0 (2):\penalty0 132--142, 2020.
\newblock \doi{10.1038/s41567-019-0733-z}.
\newblock URL \url{https://doi.org/10.1038/s41567-019-0733-z}.

\bibitem[Henriet et~al.(2020)Henriet, Beguin, Signoles, Lahaye, Browaeys,
  Reymond, and Jurczak]{Henriet2020quantum}
Lo{\"{i}}c Henriet, Lucas Beguin, Adrien Signoles, Thierry Lahaye, Antoine
  Browaeys, Georges-Olivier Reymond, and Christophe Jurczak.
\newblock Quantum computing with neutral atoms.
\newblock \emph{{Quantum}}, 4:\penalty0 327, September 2020.
\newblock ISSN 2521-327X.
\newblock \doi{10.22331/q-2020-09-21-327}.
\newblock URL \url{https://doi.org/10.22331/q-2020-09-21-327}.

\bibitem[{Morgado} and {Whitlock}(2020)]{Morgado20}
M.~{Morgado} and S.~{Whitlock}.
\newblock {Quantum simulation and computing with {R}ydberg-interacting qubits}.
\newblock \emph{arXiv e-prints}, art. arXiv:2011.03031, November 2020.

\bibitem[{Beterov}(2020)]{Beterov20}
I.~I. {Beterov}.
\newblock {Quantum Computers Based on Cold Atoms}.
\newblock \emph{Optoelectronics, Instrumentation and Data Processing},
  56\penalty0 (4):\penalty0 317--324, July 2020.
\newblock \doi{10.3103/S8756699020040020}.

\bibitem[{Wu} et~al.(2020){Wu}, {Liang}, {Tian}, {Yang}, {Chen}, {Liu}, {Khoon
  Tey}, and {You}]{Wu21}
Xiaoling {Wu}, Xinhui {Liang}, Yaoqi {Tian}, Fan {Yang}, Cheng {Chen},
  Yong-Chun {Liu}, Meng {Khoon Tey}, and Li~{You}.
\newblock {A concise review of Rydberg atom based quantum computation and
  quantum simulation}.
\newblock \emph{arXiv e-prints}, art. arXiv:2012.10614, December 2020.

\bibitem[Jaksch et~al.(2000)Jaksch, Cirac, Zoller, Rolston, C\^ot\'e, and
  Lukin]{Jaksch00}
D.~Jaksch, J.~I. Cirac, P.~Zoller, S.~L. Rolston, R.~C\^ot\'e, and M.~D. Lukin.
\newblock Fast quantum gates for neutral atoms.
\newblock \emph{Phys. Rev. Lett.}, 85:\penalty0 2208--2211, Sep 2000.
\newblock \doi{10.1103/PhysRevLett.85.2208}.
\newblock URL \url{https://link.aps.org/doi/10.1103/PhysRevLett.85.2208}.

\bibitem[Isenhower et~al.(2010)Isenhower, Urban, Zhang, Gill, Henage, Johnson,
  Walker, and Saffman]{Isenhower10}
L.~Isenhower, E.~Urban, X.~L. Zhang, A.~T. Gill, T.~Henage, T.~A. Johnson,
  T.~G. Walker, and M.~Saffman.
\newblock Demonstration of a neutral atom controlled-not quantum gate.
\newblock \emph{Phys. Rev. Lett.}, 104:\penalty0 010503, Jan 2010.
\newblock \doi{10.1103/PhysRevLett.104.010503}.
\newblock URL \url{https://link.aps.org/doi/10.1103/PhysRevLett.104.010503}.

\bibitem[Levine et~al.(2018)Levine, Keesling, Omran, Bernien, Schwartz, Zibrov,
  Endres, Greiner, Vuletić, and Lukin]{levine_high-fidelity_2018}
Harry Levine, Alexander Keesling, Ahmed Omran, Hannes Bernien, Sylvain
  Schwartz, Alexander~S. Zibrov, Manuel Endres, Markus Greiner, Vladan
  Vuletić, and Mikhail~D. Lukin.
\newblock High-{Fidelity} {Control} and {Entanglement} of {Rydberg}-{Atom}
  {Qubits}.
\newblock \emph{Phys. Rev. Lett.}, 121\penalty0 (12):\penalty0 123603,
  September 2018.
\newblock \doi{10.1103/PhysRevLett.121.123603}.
\newblock URL \url{https://link.aps.org/doi/10.1103/PhysRevLett.121.123603}.

\bibitem[Levine et~al.(2019)Levine, Keesling, Semeghini, Omran, Wang, Ebadi,
  Bernien, Greiner, Vuleti\ifmmode~\acute{c}\else \'{c}\fi{}, Pichler, and
  Lukin]{Levine19}
Harry Levine, Alexander Keesling, Giulia Semeghini, Ahmed Omran, Tout~T. Wang,
  Sepehr Ebadi, Hannes Bernien, Markus Greiner, Vladan
  Vuleti\ifmmode~\acute{c}\else \'{c}\fi{}, Hannes Pichler, and Mikhail~D.
  Lukin.
\newblock Parallel implementation of high-fidelity multiqubit gates with
  neutral atoms.
\newblock \emph{Phys. Rev. Lett.}, 123:\penalty0 170503, Oct 2019.
\newblock \doi{10.1103/PhysRevLett.123.170503}.
\newblock URL \url{https://link.aps.org/doi/10.1103/PhysRevLett.123.170503}.

\bibitem[Labuhn et~al.(2016)Labuhn, Barredo, Ravets, De~L{\'e}s{\'e}leuc,
  Macr{\`\i}, Lahaye, and Browaeys]{labuhn2016tunable}
Henning Labuhn, Daniel Barredo, Sylvain Ravets, Sylvain De~L{\'e}s{\'e}leuc,
  Tommaso Macr{\`\i}, Thierry Lahaye, and Antoine Browaeys.
\newblock Tunable two-dimensional arrays of single rydberg atoms for realizing
  quantum {I}sing models.
\newblock \emph{Nature}, 534\penalty0 (7609):\penalty0 667--670, 2016.

\bibitem[{Bernien} et~al.(2017){Bernien}, {Schwartz}, {Keesling}, {Levine},
  {Omran}, {Pichler}, {Choi}, {Zibrov}, {Endres}, {Greiner}, {Vuleti{\'c}}, and
  {Lukin}]{Bernien17}
Hannes {Bernien}, Sylvain {Schwartz}, Alexander {Keesling}, Harry {Levine},
  Ahmed {Omran}, Hannes {Pichler}, Soonwon {Choi}, Alexander~S. {Zibrov},
  Manuel {Endres}, Markus {Greiner}, Vladan {Vuleti{\'c}}, and Mikhail~D.
  {Lukin}.
\newblock {Probing many-body dynamics on a 51-atom quantum simulator}.
\newblock \emph{Nature}, 551\penalty0 (7682):\penalty0 579--584, November 2017.
\newblock \doi{10.1038/nature24622}.

\bibitem[Lienhard et~al.(2018{\natexlab{a}})Lienhard, de~L\'es\'eleuc, Barredo,
  Lahaye, Browaeys, Schuler, Henry, and L\"auchli]{Lienhard18}
Vincent Lienhard, Sylvain de~L\'es\'eleuc, Daniel Barredo, Thierry Lahaye,
  Antoine Browaeys, Michael Schuler, Louis-Paul Henry, and Andreas~M.
  L\"auchli.
\newblock Observing the space- and time-dependent growth of correlations in
  dynamically tuned synthetic ising models with antiferromagnetic interactions.
\newblock \emph{Phys. Rev. X}, 8:\penalty0 021070, Jun 2018{\natexlab{a}}.
\newblock \doi{10.1103/PhysRevX.8.021070}.
\newblock URL \url{https://link.aps.org/doi/10.1103/PhysRevX.8.021070}.

\bibitem[L{\'e}s{\'e}leuc et~al.(2019)L{\'e}s{\'e}leuc, Lienhard, Scholl,
  Barredo, Weber, Lang, Büchler, Lahaye, and
  Browaeys]{leseleuc_observation_2019}
Sylvain~de L{\'e}s{\'e}leuc, Vincent Lienhard, Pascal Scholl, Daniel Barredo,
  Sebastian Weber, Nicolai Lang, Hans~Peter Büchler, Thierry Lahaye, and
  Antoine Browaeys.
\newblock Observation of a symmetry-protected topological phase of interacting
  bosons with {Rydberg} atoms.
\newblock \emph{Science}, page eaav9105, August 2019.
\newblock ISSN 0036-8075, 1095-9203.
\newblock \doi{10.1126/science.aav9105}.
\newblock URL
  \url{https://science.sciencemag.org/content/early/2019/07/31/science.aav9105}.

\bibitem[Omran et~al.(2019)Omran, Levine, Keesling, Semeghini, Wang, Ebadi,
  Bernien, Zibrov, Pichler, Choi, Cui, Rossignolo, Rembold, Montangero,
  Calarco, Endres, Greiner, Vuletić, and Lukin]{omran_generation_2019}
Ahmed Omran, Harry Levine, Alexander Keesling, Giulia Semeghini, Tout~T. Wang,
  Sepehr Ebadi, Hannes Bernien, Alexander~S. Zibrov, Hannes Pichler, Soonwon
  Choi, Jian Cui, Marco Rossignolo, Phila Rembold, Simone Montangero, Tommaso
  Calarco, Manuel Endres, Markus Greiner, Vladan Vuletić, and Mikhail~D.
  Lukin.
\newblock Generation and manipulation of {Schr}{\textbackslash}"odinger cat
  states in {Rydberg} atom arrays.
\newblock \emph{arXiv:1905.05721 [cond-mat, physics:physics,
  physics:quant-ph]}, May 2019.
\newblock URL \url{http://arxiv.org/abs/1905.05721}.
\newblock arXiv: 1905.05721.

\bibitem[Silvério et~al.(2021)Silvério, Grijalva, Henriet, Karalekas,
  Leclerc, Shammah, and Dalyac]{Pulser}
Henrique Silvério, Sebastian Grijalva, Loïc Henriet, Peter Karalekas, Lucas
  Leclerc, Nathan Shammah, and Constantin Dalyac.
\newblock Pulser, April 2021.
\newblock URL \url{https://doi.org/10.5281/zenodo.4709401}.

\bibitem[{Parra-Rodriguez} et~al.(2020){Parra-Rodriguez}, {Lougovski},
  {Lamata}, {Solano}, and {Sanz}]{Parra-rodriguez_20}
Adrian {Parra-Rodriguez}, Pavel {Lougovski}, Lucas {Lamata}, Enrique {Solano},
  and Mikel {Sanz}.
\newblock {Digital-analog quantum computation}.
\newblock \emph{Phys. Rev. A}, 101\penalty0 (2):\penalty0 022305, February
  2020.
\newblock \doi{10.1103/PhysRevA.101.022305}.

\bibitem[Johansson et~al.(2012)Johansson, Nation, and Nori]{Johansson_2012}
J.R. Johansson, P.D. Nation, and Franco Nori.
\newblock Qu{T}i{P}: An open-source {P}ython framework for the dynamics of open
  quantum systems.
\newblock \emph{Computer Physics Communications}, 183\penalty0 (8):\penalty0
  1760–1772, Aug 2012.
\newblock ISSN 0010-4655.
\newblock \doi{10.1016/j.cpc.2012.02.021}.
\newblock URL \url{http://dx.doi.org/10.1016/j.cpc.2012.02.021}.

\bibitem[Johansson et~al.(2013)Johansson, Nation, and Nori]{Johansson_2013}
J.R. Johansson, P.D. Nation, and Franco Nori.
\newblock Qu{T}i{P} 2: A {P}ython framework for the dynamics of open quantum
  systems.
\newblock \emph{Computer Physics Communications}, 184\penalty0 (4):\penalty0
  1234–1240, Apr 2013.
\newblock ISSN 0010-4655.
\newblock \doi{10.1016/j.cpc.2012.11.019}.
\newblock URL \url{http://dx.doi.org/10.1016/j.cpc.2012.11.019}.

\bibitem[Zeng et~al.(2017)Zeng, Johnson, Smith, Rubin, Reagor, Ryan, and
  Rigetti]{Zeng_2017_Nature}
Will Zeng, Blake Johnson, Robert Smith, Nick Rubin, Matt Reagor, Colm Ryan, and
  Chad Rigetti.
\newblock First quantum computers need smart software.
\newblock \emph{Nature News}, 549\penalty0 (7671):\penalty0 149, 2017.

\bibitem[Karalekas et~al.(2020)Karalekas, Tezak, Peterson, Ryan, da~Silva, and
  Smith]{Karalekas_2020}
Peter~J Karalekas, Nikolas~A Tezak, Eric~C Peterson, Colm~A Ryan, Marcus~P
  da~Silva, and Robert~S Smith.
\newblock A quantum-classical cloud platform optimized for variational hybrid
  algorithms.
\newblock \emph{Quantum Science and Technology}, 5\penalty0 (2):\penalty0
  024003, Apr 2020.
\newblock \doi{10.1088/2058-9565/ab7559}.
\newblock URL \url{https://doi.org/10.1088%2F2058-9565%2Fab7559}.

\bibitem[McClean et~al.(2016)McClean, Romero, Babbush, and
  Aspuru-Guzik]{McClean_2016_NJP}
Jarrod~R McClean, Jonathan Romero, Ryan Babbush, and Al{\'a}n Aspuru-Guzik.
\newblock The theory of variational hybrid quantum-classical algorithms.
\newblock \emph{New Journal of Physics}, 18\penalty0 (2):\penalty0 023023,
  2016.

\bibitem[Bharti et~al.(2021)Bharti, Cervera-Lierta, Kyaw, Haug, Alperin-Lea,
  Anand, Degroote, Heimonen, Kottmann, Menke, Mok, Sim, Kwek, and
  Aspuru-Guzik]{Bharti_2021}
Kishor Bharti, Alba Cervera-Lierta, Thi~Ha Kyaw, Tobias Haug, Sumner
  Alperin-Lea, Abhinav Anand, Matthias Degroote, Hermanni Heimonen, Jakob~S.
  Kottmann, Tim Menke, Wai-Keong Mok, Sukin Sim, Leong-Chuan Kwek, and Alán
  Aspuru-Guzik.
\newblock Noisy intermediate-scale quantum ({N}{I}{S}{Q}) algorithms, 2021.

\bibitem[Endo et~al.(2021)Endo, Cai, Benjamin, and Yuan]{Endo_2021}
Suguru Endo, Zhenyu Cai, Simon~C. Benjamin, and Xiao Yuan.
\newblock Hybrid quantum-classical algorithms and quantum error mitigation.
\newblock \emph{Journal of the Physical Society of Japan}, 90\penalty0
  (3):\penalty0 032001, Mar 2021.
\newblock ISSN 1347-4073.
\newblock \doi{10.7566/jpsj.90.032001}.
\newblock URL \url{http://dx.doi.org/10.7566/JPSJ.90.032001}.

\bibitem[Alexander et~al.(2020)Alexander, Kanazawa, Egger, Capelluto, Wood,
  Javadi-Abhari, and C~McKay]{Alexander_2020_QST}
Thomas Alexander, Naoki Kanazawa, Daniel~J Egger, Lauren Capelluto,
  Christopher~J Wood, Ali Javadi-Abhari, and David C~McKay.
\newblock Qiskit pulse: programming quantum computers through the cloud with
  pulses.
\newblock \emph{Quantum Science and Technology}, 5\penalty0 (4):\penalty0
  044006, Aug 2020.
\newblock ISSN 2058-9565.
\newblock \doi{10.1088/2058-9565/aba404}.
\newblock URL \url{http://dx.doi.org/10.1088/2058-9565/aba404}.

\bibitem[Ball et~al.(2020)Ball, Biercuk, Carvalho, Chen, Hush, Castro, Li,
  Liebermann, Slatyer, Edmunds, Frey, Hempel, and Milne]{Ball_2020}
Harrison Ball, Michael~J. Biercuk, Andre Carvalho, Jiayin Chen, Michael Hush,
  Leonardo A.~De Castro, Li~Li, Per~J. Liebermann, Harry~J. Slatyer, Claire
  Edmunds, Virginia Frey, Cornelius Hempel, and Alistair Milne.
\newblock Software tools for quantum control: Improving quantum computer
  performance through noise and error suppression, 2020.

\bibitem[Wittler et~al.(2021)Wittler, Roy, Pack, Werninghaus, Roy, Egger,
  Filipp, Wilhelm, and Machnes]{Wittler_2021}
Nicolas Wittler, Federico Roy, Kevin Pack, Max Werninghaus, Anurag~Saha Roy,
  Daniel~J. Egger, Stefan Filipp, Frank~K. Wilhelm, and Shai Machnes.
\newblock Integrated tool set for control, calibration, and characterization of
  quantum devices applied to superconducting qubits.
\newblock \emph{Physical Review Applied}, 15\penalty0 (3), Mar 2021.
\newblock ISSN 2331-7019.
\newblock \doi{10.1103/physrevapplied.15.034080}.
\newblock URL \url{http://dx.doi.org/10.1103/PhysRevApplied.15.034080}.

\bibitem[Winer et~al.(2021)Winer, Romach, Bade, Mitnikov, Shani, Israeli,
  Cohen, Abend, and Ella]{QUA-libs}
Gal Winer, Yoav Romach, Satya Bade, Ilan Mitnikov, Tal Shani, Dor Israeli,
  Yonatan Cohen, Uri Abend, and Lior Ella.
\newblock qua-libs: Open source libraries in qua, May 2021.
\newblock URL \url{https://doi.org/10.5281/zenodo.4769470}.

\bibitem[Li et~al.(2021)Li, Ahmed, Saraogi, Lambert, Nori, Pitchford, and
  Shammah]{li2021qutip-qip}
Boxi Li, Shahnawaz Ahmed, Sidhant Saraogi, Neill Lambert, Franco Nori,
  Alexander Pitchford, and Nathan Shammah.
\newblock Pulse-level noisy quantum circuits with qutip.
\newblock \emph{arXiv preprint arXiv:2105.09902}, 2021.

\bibitem[Goerz et~al.(2019)Goerz, Basilewitsch, Gago-Encinas, Krauss, Horn,
  Reich, and Koch]{Goerz_2019_SciPost}
Michael~H. Goerz, Daniel Basilewitsch, Fernando Gago-Encinas, Matthias~G.
  Krauss, Karl~P. Horn, Daniel~M. Reich, and Christiane~P. Koch.
\newblock {Krotov: A Python implementation of {K}rotov's method for quantum
  optimal control}.
\newblock \emph{SciPost Phys.}, 7:\penalty0 80, 2019.
\newblock \doi{10.21468/SciPostPhys.7.6.080}.
\newblock URL \url{https://scipost.org/10.21468/SciPostPhys.7.6.080}.

\bibitem[Preskill(2018)]{Preskill_2018_Quantum}
John Preskill.
\newblock Quantum computing in the nisq era and beyond.
\newblock \emph{Quantum}, 2:\penalty0 79, Aug 2018.
\newblock ISSN 2521-327X.
\newblock \doi{10.22331/q-2018-08-06-79}.
\newblock URL \url{http://dx.doi.org/10.22331/q-2018-08-06-79}.

\bibitem[Nogrette et~al.(2014)Nogrette, Labuhn, Ravets, Barredo, B\'eguin,
  Vernier, Lahaye, and Browaeys]{Nogrette14}
F.~Nogrette, H.~Labuhn, S.~Ravets, D.~Barredo, L.~B\'eguin, A.~Vernier,
  T.~Lahaye, and A.~Browaeys.
\newblock Single-atom trapping in holographic 2d arrays of microtraps with
  arbitrary geometries.
\newblock \emph{Phys. Rev. X}, 4:\penalty0 021034, May 2014.
\newblock \doi{10.1103/PhysRevX.4.021034}.
\newblock URL \url{https://link.aps.org/doi/10.1103/PhysRevX.4.021034}.

\bibitem[Schau{\ss} et~al.(2015)Schau{\ss}, Zeiher, Fukuhara, Hild, Cheneau,
  Macr{\`\i}, Pohl, Bloch, and Gro{\ss}]{schauss2015crystallization}
Peter Schau{\ss}, Johannes Zeiher, Takeshi Fukuhara, Sebastian Hild, Marc
  Cheneau, Tommaso Macr{\`\i}, Thomas Pohl, Immanuel Bloch, and Christian
  Gro{\ss}.
\newblock Crystallization in {I}sing quantum magnets.
\newblock \emph{Science}, 347\penalty0 (6229):\penalty0 1455--1458, 2015.

\bibitem[de~L\'es\'eleuc et~al.(2018)de~L\'es\'eleuc, Weber, Lienhard, Barredo,
  B\"uchler, Lahaye, and Browaeys]{leseleuc2018accurate}
Sylvain de~L\'es\'eleuc, Sebastian Weber, Vincent Lienhard, Daniel Barredo,
  Hans~Peter B\"uchler, Thierry Lahaye, and Antoine Browaeys.
\newblock Accurate mapping of multilevel {R}ydberg atoms on interacting
  spin-$1/2$ particles for the quantum simulation of {I}sing models.
\newblock \emph{Phys. Rev. Lett.}, 120:\penalty0 113602, Mar 2018.
\newblock \doi{10.1103/PhysRevLett.120.113602}.
\newblock URL \url{https://link.aps.org/doi/10.1103/PhysRevLett.120.113602}.

\bibitem[Barredo et~al.(2015)Barredo, Labuhn, Ravets, Lahaye, Browaeys, and
  Adams]{barredo2015coherent}
Daniel Barredo, Henning Labuhn, Sylvain Ravets, Thierry Lahaye, Antoine
  Browaeys, and Charles~S. Adams.
\newblock Coherent excitation transfer in a spin chain of three {R}ydberg
  atoms.
\newblock \emph{Phys. Rev. Lett.}, 114:\penalty0 113002, Mar 2015.
\newblock \doi{10.1103/PhysRevLett.114.113002}.
\newblock URL \url{https://link.aps.org/doi/10.1103/PhysRevLett.114.113002}.

\bibitem[Orioli et~al.(2018)Orioli, Signoles, Wildhagen, G\"unter, Berges,
  Whitlock, and Weidem\"uller]{orioli2018relaxation}
A.~Pi\~neiro Orioli, A.~Signoles, H.~Wildhagen, G.~G\"unter, J.~Berges,
  S.~Whitlock, and M.~Weidem\"uller.
\newblock Relaxation of an isolated dipolar-interacting {R}ydberg quantum spin
  system.
\newblock \emph{Phys. Rev. Lett.}, 120:\penalty0 063601, Feb 2018.
\newblock \doi{10.1103/PhysRevLett.120.063601}.
\newblock URL \url{https://link.aps.org/doi/10.1103/PhysRevLett.120.063601}.

\bibitem[{Scholl} et~al.(2021){Scholl}, {Williams}, {Bornet}, {Wallner},
  {Barredo}, {Lahaye}, {Browaeys}, {Henriet}, {Signoles}, {Hainaut}, {Franz},
  {Geier}, {Tebben}, {Salzinger}, {Z{\"u}rn}, and
  {Weidem{\"u}ller}]{Scholl2021}
P.~{Scholl}, H.~J. {Williams}, G.~{Bornet}, F.~{Wallner}, D.~{Barredo},
  T.~{Lahaye}, A.~{Browaeys}, L.~{Henriet}, A.~{Signoles}, C.~{Hainaut},
  T.~{Franz}, S.~{Geier}, A.~{Tebben}, A.~{Salzinger}, G.~{Z{\"u}rn}, and
  M.~{Weidem{\"u}ller}.
\newblock {Microwave-engineering of programmable XXZ Hamiltonians in arrays of
  Rydberg atoms}.
\newblock \emph{arXiv e-prints}, art. arXiv:2107.14459, July 2021.

\bibitem[Bluvstein et~al.(2021)Bluvstein, Omran, Levine, Keesling, Semeghini,
  Ebadi, Wang, Michailidis, Maskara, Ho, Choi, Serbyn, Greiner, Vuleti{\'c},
  and Lukin]{Bluvstein21}
D.~Bluvstein, A.~Omran, H.~Levine, A.~Keesling, G.~Semeghini, S.~Ebadi, T.~T.
  Wang, A.~A. Michailidis, N.~Maskara, W.~W. Ho, S.~Choi, M.~Serbyn,
  M.~Greiner, V.~Vuleti{\'c}, and M.~D. Lukin.
\newblock Controlling quantum many-body dynamics in driven rydberg atom arrays.
\newblock \emph{Science}, 2021.
\newblock ISSN 0036-8075.
\newblock \doi{10.1126/science.abg2530}.
\newblock URL
  \url{https://science.sciencemag.org/content/early/2021/02/24/science.abg2530}.

\bibitem[Scholl et~al.(2020)Scholl, Schuler, Williams, Eberharter, Barredo,
  Schymik, Lienhard, Henry, Lang, Lahaye, et~al.]{scholl2020programmable}
Pascal Scholl, Michael Schuler, Hannah~J Williams, Alexander~A Eberharter,
  Daniel Barredo, Kai-Niklas Schymik, Vincent Lienhard, Louis-Paul Henry,
  Thomas~C Lang, Thierry Lahaye, et~al.
\newblock Programmable quantum simulation of 2d antiferromagnets with hundreds
  of rydberg atoms.
\newblock \emph{arXiv preprint arXiv:2012.12268}, 2020.

\bibitem[Ebadi et~al.(2020)Ebadi, Wang, Levine, Keesling, Semeghini, Omran,
  Bluvstein, Samajdar, Pichler, Ho, et~al.]{ebadi2020quantum}
Sepehr Ebadi, Tout~T Wang, Harry Levine, Alexander Keesling, Giulia Semeghini,
  Ahmed Omran, Dolev Bluvstein, Rhine Samajdar, Hannes Pichler, Wen~Wei Ho,
  et~al.
\newblock Quantum phases of matter on a 256-atom programmable quantum
  simulator.
\newblock \emph{arXiv preprint arXiv:2012.12281}, 2020.

\bibitem[Pichler et~al.(2018)Pichler, Wang, Zhou, Choi, and
  Lukin]{pichler2018quantum}
Hannes Pichler, Sheng-Tao Wang, Leo Zhou, Soonwon Choi, and Mikhail~D Lukin.
\newblock Quantum optimization for maximum independent set using rydberg atom
  arrays.
\newblock \emph{arXiv preprint arXiv:1808.10816}, 2018.

\bibitem[Henriet(2020)]{henriet2020robustness}
Lo{\"\i}c Henriet.
\newblock Robustness to spontaneous emission of a variational quantum
  algorithm.
\newblock \emph{Physical Review A}, 101\penalty0 (1):\penalty0 012335, 2020.

\bibitem[Dalyac et~al.(2020)Dalyac, Henriet, Jeandel, Lechner, Perdrix,
  Porcheron, and Veshchezerova]{dalyac2020qualifying}
Constantin Dalyac, Lo{\"\i}c Henriet, Emmanuel Jeandel, Wolfgang Lechner, Simon
  Perdrix, Marc Porcheron, and Margarita Veshchezerova.
\newblock Qualifying quantum approaches for hard industrial optimization
  problems. a case study in the field of smart-charging of electric vehicles.
\newblock \emph{arXiv preprint arXiv:2012.14859}, 2020.

\bibitem[McKay et~al.(2017)McKay, Wood, Sheldon, Chow, and
  Gambetta]{IBM_VZ-gates}
David~C. McKay, Christopher~J. Wood, Sarah Sheldon, Jerry~M. Chow, and Jay~M.
  Gambetta.
\newblock Efficient $z$ gates for quantum computing.
\newblock \emph{Phys. Rev. A}, 96:\penalty0 022330, Aug 2017.
\newblock \doi{10.1103/PhysRevA.96.022330}.
\newblock URL \url{https://link.aps.org/doi/10.1103/PhysRevA.96.022330}.

\bibitem[Fuhrmanek et~al.(2011)Fuhrmanek, Bourgain, Sortais, and
  Browaeys]{Fuhrmanek11}
A.~Fuhrmanek, R.~Bourgain, Y.~R.~P. Sortais, and A.~Browaeys.
\newblock Free-space lossless state detection of a single trapped atom.
\newblock \emph{Phys. Rev. Lett.}, 106:\penalty0 133003, Mar 2011.
\newblock \doi{10.1103/PhysRevLett.106.133003}.
\newblock URL \url{https://link.aps.org/doi/10.1103/PhysRevLett.106.133003}.

\bibitem[Lienhard et~al.(2018{\natexlab{b}})Lienhard, de~L{\'e}s{\'e}leuc,
  Barredo, Lahaye, Browaeys, Schuler, Henry, and
  L{\"a}uchli]{lienhard2018observing}
Vincent Lienhard, Sylvain de~L{\'e}s{\'e}leuc, Daniel Barredo, Thierry Lahaye,
  Antoine Browaeys, Michael Schuler, Louis-Paul Henry, and Andreas~M
  L{\"a}uchli.
\newblock Observing the space-and time-dependent growth of correlations in
  dynamically tuned synthetic ising models with antiferromagnetic interactions.
\newblock \emph{Physical Review X}, 8\penalty0 (2):\penalty0 021070,
  2018{\natexlab{b}}.

\bibitem[{Farhi} et~al.(2014){Farhi}, {Goldstone}, and {Gutmann}]{Farhi14}
Edward {Farhi}, Jeffrey {Goldstone}, and Sam {Gutmann}.
\newblock {A Quantum Approximate Optimization Algorithm}.
\newblock \emph{arXiv e-prints}, art. arXiv:1411.4028, Nov 2014.

\bibitem[{Schymik} et~al.(2020){Schymik}, {Lienhard}, {Barredo}, {Scholl},
  {Williams}, {Browaeys}, and {Lahaye}]{Schymik20}
Kai-Niklas {Schymik}, Vincent {Lienhard}, Daniel {Barredo}, Pascal {Scholl},
  Hannah {Williams}, Antoine {Browaeys}, and Thierry {Lahaye}.
\newblock {Enhanced atom-by-atom assembly of arbitrary tweezer arrays}.
\newblock \emph{Phys. Rev. A}, 102\penalty0 (6):\penalty0 063107, December
  2020.
\newblock \doi{10.1103/PhysRevA.102.063107}.

\bibitem[{Pichler} et~al.(2018){Pichler}, {Wang}, {Zhou}, {Choi}, and
  {Lukin}]{Pichler18}
Hannes {Pichler}, Sheng-Tao {Wang}, Leo {Zhou}, Soonwon {Choi}, and Mikhail~D.
  {Lukin}.
\newblock {Quantum Optimization for Maximum Independent Set Using Rydberg Atom
  Arrays}.
\newblock \emph{arXiv e-prints}, art. arXiv:1808.10816, Aug 2018.

\bibitem[Zhou et~al.(2020)Zhou, Wang, Choi, Pichler, and Lukin]{zhou20}
Leo Zhou, Sheng-Tao Wang, Soonwon Choi, Hannes Pichler, and Mikhail~D. Lukin.
\newblock Quantum approximate optimization algorithm: Performance, mechanism,
  and implementation on near-term devices.
\newblock \emph{Phys. Rev. X}, 10:\penalty0 021067, Jun 2020.
\newblock \doi{10.1103/PhysRevX.10.021067}.
\newblock URL \url{https://link.aps.org/doi/10.1103/PhysRevX.10.021067}.

\bibitem[Henry et~al.(2021)Henry, Thabet, Dalyac, and Henriet]{Henry21}
Louis-Paul Henry, Slimane Thabet, Constantin Dalyac, and Lo\"{\i}c Henriet.
\newblock Quantum evolution kernel: Machine learning on graphs with
  programmable arrays of qubits.
\newblock \emph{Phys. Rev. A}, 104:\penalty0 032416, Sep 2021.
\newblock \doi{10.1103/PhysRevA.104.032416}.
\newblock URL \url{https://link.aps.org/doi/10.1103/PhysRevA.104.032416}.

\bibitem[Developers(2021)]{cirq_developers_2021_4586899}
Cirq Developers.
\newblock Cirq, March 2021.
\newblock URL \url{https://doi.org/10.5281/zenodo.4586899}.
\newblock {See full list of authors on Github:
  \href{https://github.com/quantumlib/Cirq/graphs/contributors}{Cirq/graphs/contributors}}.

\bibitem[Developers(2019)]{Qiskit}
Qiskit Developers.
\newblock Qiskit: An open-source framework for quantum computing, 2019.

\bibitem[{Bergholm} et~al.(2018){Bergholm}, {Izaac}, {Schuld}, {Gogolin},
  {Sohaib Alam}, {Ahmed}, {Arrazola}, {Blank}, {Delgado}, {Jahangiri},
  {McKiernan}, {Meyer}, {Niu}, {Sz{\'a}va}, and {Killoran}]{Pennylane}
Ville {Bergholm}, Josh {Izaac}, Maria {Schuld}, Christian {Gogolin}, M.~{Sohaib
  Alam}, Shahnawaz {Ahmed}, Juan~Miguel {Arrazola}, Carsten {Blank}, Alain
  {Delgado}, Soran {Jahangiri}, Keri {McKiernan}, Johannes~Jakob {Meyer}, Zeyue
  {Niu}, Antal {Sz{\'a}va}, and Nathan {Killoran}.
\newblock {PennyLane: Automatic differentiation of hybrid quantum-classical
  computations}.
\newblock \emph{arXiv e-prints}, art. arXiv:1811.04968, November 2018.

\bibitem[{Sivarajah} et~al.(2021){Sivarajah}, {Dilkes}, {Cowtan}, {Simmons},
  {Edgington}, and {Duncan}]{tket}
Seyon {Sivarajah}, Silas {Dilkes}, Alexander {Cowtan}, Will {Simmons}, Alec
  {Edgington}, and Ross {Duncan}.
\newblock {t|ket⟩: a retargetable compiler for NISQ devices}.
\newblock \emph{Quantum Science and Technology}, 6\penalty0 (1):\penalty0
  014003, January 2021.
\newblock \doi{10.1088/2058-9565/ab8e92}.

\bibitem[{de L{\'e}s{\'e}leuc} et~al.(2018){de L{\'e}s{\'e}leuc}, {Barredo},
  {Lienhard}, {Browaeys}, and {Lahaye}]{sylvain18}
Sylvain {de L{\'e}s{\'e}leuc}, Daniel {Barredo}, Vincent {Lienhard}, Antoine
  {Browaeys}, and Thierry {Lahaye}.
\newblock {Analysis of imperfections in the coherent optical excitation of
  single atoms to Rydberg states}.
\newblock \emph{Phys. Rev. A}, 97\penalty0 (5):\penalty0 053803, May 2018.
\newblock \doi{10.1103/PhysRevA.97.053803}.

\end{thebibliography}
\end{document}